\documentclass[acmsmall,screen, nonacm]{acmart}

\setcopyright{acmlicensed}
\copyrightyear{2025}
\acmYear{2025}
\acmDOI{XXXXXXX.XXXXXXX}
\acmConference[DEBS'25]{19th ACM Int. Conf. Distributed and Event-based Systems}{June 10--13,
  2025}{Gothenburg}
\acmISBN{978-1-4503-XXXX-X/2025/08}

\usepackage[utf8]{inputenc}
\usepackage[T1]{fontenc}
\newcommand{\newbowtie}{\mathrel{\ooalign{$\triangleright$\,\cr\,$\triangleleft$}}}
\usepackage{soul,color}
\usepackage{xcolor,colortbl}
\newtheorem{definition}{}
\usepackage{hyperref}
\usepackage{tikz}
\usetikzlibrary{positioning,chains,fit,shapes,calc}
\usetikzlibrary{shapes.geometric, arrows, calc, arrows.meta}
\usepackage{tkz-euclide}
\usepackage{lipsum}
\usepackage{placeins}
\usepackage{bm}
\usepackage[utf8]{inputenc}
\usepackage[linesnumbered,ruled,vlined]{algorithm2e}
\usepackage{subfigure}
\usepackage[edges]{forest}
\usepackage{enumitem}
\usepackage{balance}
\usepackage[linesnumbered,ruled,vlined]{algorithm2e}
\usepackage{mathtools}
\DeclarePairedDelimiter\ceil{\lceil}{\rceil}

\SetCommentSty{mycommfont}
\SetKwInput{KwInput}{Input}               
\SetKwInput{KwOutput}{Output}  
\SetKwProg{Fn}{Function}{}{}
\usepackage{mathtools}
\DeclarePairedDelimiter{\floor}{\lfloor}{\rfloor}
\usepackage{makecell}

\definecolor{c1}{rgb}{0.358, 0.188, 0.478}
\definecolor{c2}{rgb}{0.558, 0.188, 0.478}
\definecolor{c3}{rgb}{0.668, 0.188, 0.478}
\definecolor{c4}{rgb}{0.768, 0.188, 0.478}
\definecolor{c5}{rgb}{0.888, 0.188, 0.478}
\definecolor{darkorange}{rgb}{0.8, 0.4, 0.0}
\definecolor{mustard0}{rgb}{0.9, 0.8, 0.3}
\definecolor{mustard}{rgb}{0.9, 0.5, 0.4}
\definecolor{cherry}{rgb}{0.8, 0.3, 0.5}
\definecolor{cherrybrown}{rgb}{0.8, 0.3, 0.2}
\definecolor{lightpurple}{rgb}{.4, .7, .4}
\definecolor{green1}{rgb}{.1, .7, .7}

\begin{document}

\title{Transient Concepts in Streaming Graphs}

\author{Aida Sheshbolouki}
\email{aida.sheshbolouki@uwaterloo.ca}
\orcid{0000-0001-5725-1781}

\author{M. Tamer {\"O}zsu}
\affiliation{
  \institution{University of Waterloo}
  \country{Canada}
  }
\email{tamer.ozsu@uwaterloo.ca}
\orcid{0000-0002-8126-1717}


\begin{abstract}
Concept Drift (CD) occurs when a change in a hidden context can induce changes in a target concept. CD is a natural phenomenon in non-stationary settings such as data streams.  Understanding, detection, and adaptation to CD in streaming data is (i) vital for effective and efficient analytics as reliable output depends on adaptation to fresh input, (ii) challenging as it requires efficient operations as well as effective performance evaluations, and (iii) impactful as it applies to a variety of use cases and is a crucial initial step for data management systems. Current works are mostly focused on passive CD detection as part of supervised adaptation, on independently generated data instances or graph snapshots, on target concepts as a function of data labels, on static data management, and on specific temporal order of data record. These methods do not always work.  We revisit CD for the streaming graphs setting and introduce two first-of-its-kind frameworks \emph{SGDD} and \emph{SGDP} for streaming graph CD detection and prediction. Both frameworks discern the change of generative source. \emph{SGDD} detects the CDs due to the changes of generative parameters with significant delays such that it is difficult to evaluate the performance, while \emph{SGDP} predicts these CDs between 7374 to 0.19 milliseconds ahead of their occurrence, without accessing the payloads of data records. 
\end{abstract}

\ccsdesc[500]{Computer systems organization~Reliability}
\ccsdesc[500]{Information systems~Data stream mining}

\keywords{Concept Drift, Streaming Graphs}

\maketitle

\section{Introduction}

Systems that run a data-driven decision making task (e.g., training/testing a learner model or answering a user-specified query), generate unreliable outputs when the input data or the decision factors are new, temporal, incomplete, or manipulated if the task does not recognize and manage it. A main cause identified for this problem is \textit{concept drift} (CD)~\cite{lu2019learning}. CD is a phenomenon that occurs when ``changes in hidden context can induce  more or less radical changes in target concept''~\cite{widmer1996learning}. Hidden context refers to insufficient, incomplete, or unobservable information about input data~\cite{elwell2011incremental}. Target concept refers to known and/or observable information that have direct impact on the task's output.  For instance, change of user opinions or customer relationship management affects the rating patterns; the data arrival from a certain location or type of users gets interrupted or changes, and this affects the distribution of arrivals;  a mental health issue affects the driving patterns and generation of driving tickets or traffic control data; a biological function fluctuates the blood sugar level or heart beat rate; external factors deteriorate a wound tissue and the oxygen level drops and the temperature increases in the wound area.  \begin{figure}
    \centering
    \includegraphics[width=\linewidth]{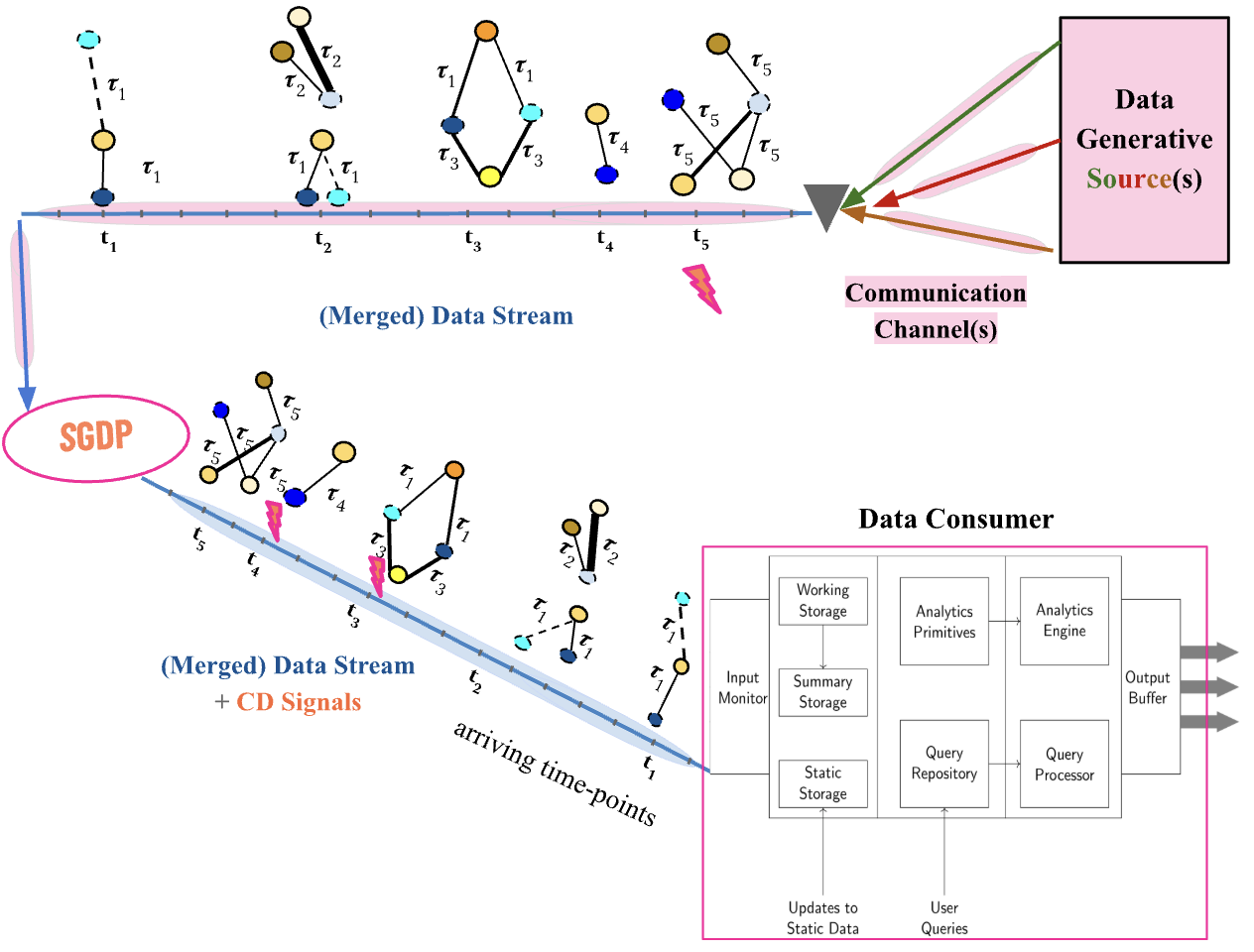}
    \caption{Our CD Management scheme.}
    \label{fig:CDManagementFramework2}
\end{figure}

In non-stationary settings such as that of streaming data, CD is natural~\cite{kifer2004detecting, moreno2012unifying}. 
CD management covers three aspects~\cite{lu2019learning}: CD detection to identify changes to characterize and quantify the drift; CD understanding to describe the drift event by providing information about the time, severity, and/or the contributing factors of drift; and CD adaptation to update a downstream task. Prompt CD management in streaming settings is important for generating relevant, reliable, and effective outputs. CD detection and understanding, our focus in this paper, benefit (i) development of accurate generative data models which are expected to (not) preserve concepts, (ii)  generalization of analytics, (iii) designing algorithms (e.g., network protocols for error control and estimating the time-to-live of routing packets), and (iv) anomaly detections for ``real-time monitoring or control of some automated activity'', ``organizing and personalizing information'', and ``characterizing health, well-being, or a state of humans, economics, or entities''~\cite{vzliobaite2016overview}.

In streaming data model, the streaming rate is highly dynamic and incurs simultaneous arrival of several bursts of data records, each generated at a time point. In applications where the payloads of data records are interconnected, it is helpful to model the data as a graph. The combination of graph data and streaming leads to \textit{streaming graphs}. Current CD management techniques do not consider this setting and generally incorporate a number of assumptions and design choices which do not always work properly due to the following. 

    \textbf{a)} The data is  a sequence of totally ordered graph snapshots with vertex attributes and task labels~\cite{zambon2018concept,paudel2020approach} and/or drift detection is integrated in online supervised systems, which means drifts are detected after one or more performance drops are observed in the data consumer task~\cite{wares2019data,halstead2022analyzing, zhou2023multi, zhou2024dynamic, gama2004learning,liu2017fuzzy,xu2017dynamic,frias2014online}. The performance drop is  associated to some sort of unidentified distribution change without proper justification~\cite{ zhou2023multi, zhou2024dynamic}. Another supervised CD detection method uses CD labels which differ from the labels of the consumer (task labels). These supervised settings are not always helpful:
        \begin{itemize}
        \item
        [\boldsymbol{$*$}] The drift signals cannot be used for other target tasks and the detection process should be repeated in a system performing multiple tasks on the same data (e.g., a large knowledge graph). The tasks may not require or use the same concepts and data labels.
        \item 
        [\boldsymbol{$*$}] The task/CD labels and graph attributes are not always available due to privacy concerns, the cost of feature engineering, the cost of storing processed data, and the unknown nature of CD events, etc.
        \item[\boldsymbol{$*$}] While performance testing for CD detection runs, newly arrived data is processed or kept until the task is possibly updated. This delays the outputs and the CD signals can only help with unprocessed data. Correcting the previous outputs and re-processing is expensive and not always feasible.
        \item[\boldsymbol{$*$}] The performance drop should not always be related to CD; e.g., the performance of a data model can decrease because the transductive model cannot be applied to new data, because the data is incomplete with missing representative samples or missing record elements, because the data is not clean, because the data is manipulated, etc.  
        \item[\boldsymbol{$*$}] A performance drop caused by CD can be due to either a change in a data distribution (virtual CD) or a change in the relation between data and supervision labels (real CD). It is important to discern the CD type, which requires further computes.
        \item[\boldsymbol{$*$}] The performance of a data model does not necessarily drop or detect a CD. E.g., the model can be robust to CD, boosted by a CD , or miss a slow CD. 
        \item[\boldsymbol{$*$}] Frequent performance checks can be wasteful when CDs are not frequent or abrupt.
        \end{itemize}
        
    \textbf{b)} The methods operate on streaming data records which are independently generated~\cite{wares2019data,zhou2023multi, zhou2024dynamic, kifer2004detecting} and compare the underlying distribution of sequential sets of data~\cite{halstead2022analyzing,bontempelli2022human}. Analyzing the payload of graph data records without considering the interconnectivity among them and their generation timestamps is not sound and complete.
        \begin{itemize}
        \item
        [\boldsymbol{$*$}] Using i.i.d. assumptions for graph edges which are not independently generated is misleading. 
        \item
        [\boldsymbol{$*$}] The data records may arrive out-of-order or repeatedly; Using time-based sliding windows based on just arrival times or just generation timestamps would incur processing with incorrect temporal information.   
        \end{itemize}

        \textbf{c)} The data is a sequence of windowed graphs or CD is detected using static parameters (e.g., a drift threshold, prototype baselines, window size, and slide size)~\cite{halstead2022analyzing,zhou2023multi, zhou2024dynamic, nishida2007detecting, yao2016detecting, babcock2002models, yao2016detecting,zambon2018concept,paudel2020approach}. Analyzing streaming edges with fixed parameters is not effective enough.
        \begin{itemize}
            \item[\boldsymbol{$*$}] Fixed drift thresholds cannot adapt to new concepts.
            \item[\boldsymbol{$*$}] Fixing the window (graph snapshot) sizes to static numbers is neither efficient nor effective with the highly dynamic streaming rates. 
            \item[\boldsymbol{$*$}] An edge stream can be a mix of several streams generated by several sources and differentiating streams could incur additional costs (e.g., to separate the arrivals at one processing node). Graph streams commonly do not capture this heterogeneity. 
            \item[\boldsymbol{$*$}] Evaluating the performance of batched analyses on graph streams is challenging. Accuracy can be influenced by the batch size. When several changes occur sequentially and the detection algorithm relies on historical information,  a large batch and consequently a large system state, delays the detection outcomes, exhausts the memory, and the late detections can be viewed as missed/incorrect  detections. 
        \end{itemize}

 We revisit the transient concepts in streaming graphs to solve this problem: \emph{Given the unbounded sub-sequence of streaming graph records, which are captured after a certain start point and partially ordered by their arrival time, how to signal a CD, while providing descriptive information about the drift without using supervision data labels.} We focus on the generative source as the hidden context (i.e., we signal the changes in the generative source(s) of streaming graph records) and do not consider factors such as merging, and sampling. 

we define CD in streaming graphs as a change in a characteristic data pattern. We reduce the problem of CD signaling in streaming graphs to change detection in time-series of major data patterns. We choose data patterns that reflect the generative patterns of butterflies ((2,2)-bicliques) since (i) the streaming graph record (SGR)s in many applications capture the interactions that naturally occur in a bipartite mode, (ii) all complex networks have an underlying bipartite structure driving the topological structure of the unipartite version \cite{ guillaume2004bipartite}, (iii) butterflies are characteristic substructures in bipartite streaming graphs~\cite{sheshbolouki2022sgrapp, Sheshbolouki:2023aa}, and (iv) butterflies can be listed incrementally without re-examining their existence.\footnote{Maximal subgraphs including several butterflies such as k-bitruss~\cite{wang2020efficient}, k-wing~\cite{sariyuce2018peeling}, and $ s(\alpha,\beta)_\tau-$core~\cite{he2021exploring}) require dynamic maintenance.} 

We introduce two frameworks: 
\emph{\textcolor{c1}{S}\textcolor{c2}{G}\textcolor{c3}{D}\textcolor{c4}{D}},  for \textcolor{c1}{s}treaming \textcolor{c2}{g}raph \textcolor{c3}{d}rift \textcolor{c4}{d}etection via tracking the interconnectivity of butterflies, which serves as a baseline for an advanced framework \emph{\textcolor{c1}{S}\textcolor{c2}{G}\textcolor{c3}{D}\textcolor{c4}{P}}, for \textcolor{c1}{s}treaming \textcolor{c2}{g}raph \textcolor{c3}{d}rift \textcolor{c4}{p}rediction via tracking the burstiness of the streaming edges.
Both frameworks support any downstream analytics (supervised or unsupervised), explain the time and location of drifts,
are unsupervised, adapt to the streaming rate, and do not require any input parameter. \emph{SGDP} advances \emph{SGDD} as it predicts CDs without utilizing data payloads (lightweight computations, privacy-preserving, and applicable to any data stream).



We demonstrate \emph{SGDP} with the following motivating example. Figure~\ref{fig:CDManagementFramework2} shows a sequence of SGRs (each with a generation timestamp $\tau_i$ and a non-empty payload) generated by one or more sources; Each SGR arrives at the data consumer at a time point $t_j$. The data records with the generation timestamp $\tau_5$ and arrival time point $t_5$ are impacted by CD (a change in a hidden context e.g., an error in the transmission channel, a man-in-the-middle attack, a change of generative process or its parameters). \emph{SGDP} receives the SGRs and 
at $t_3$ and $t_4$, CD signals are streamed out (before the arrival of impacted data records at $t_5$). 
\emph{SGDP} monitors data patterns \textbf{without accessing the SGR payloads}. Moreover, it does not change the distribution of data arriving at the consumer system; it can and should be implemented in the input buffer of the consumer system for comprehensive CD checks or as a middleware exposed to further CDs. The CD signals can be incorporated in networking protocols (e.g., flipping a bit in the packet header with encryptions, sending a separate control packet, or watermark data annotation) -- the specific implementation and notification methods are beyond the scope of this paper. The data consumer ingests the data records and the CD signals, and activates appropriate CD adaptation mechanisms  (e.g., regulating the ingestion, data cleaning, updating the operators/storage/primitives/results, pushing forward the CD alerts as detected anomalies, etc).


Section~\ref{sec:definitions} is a dictionary of terminology and notations. Section~\ref{sec:relatedworks} reviews the methods on CD management. Sections~\ref{sec:sGradP} introduces  \emph{SGDP}. Section~\ref{sec:evaluations} introduces \emph{SGDD} (details in Appendices~\ref{appendixA}) and includes the performance evaluations. Section~\ref{sec:conclusion} concludes the paper.

\section{Dictionary}\label{sec:definitions}

\begin{definition}[Streaming Record (SR), $r_m$]\label{def:streamrecord}
 A 2-tuple $r_m$=$\langle p,\tau \rangle$ where $p$ defines the payload of the record and $\tau$ is its event (application) timestamp  assigned by the data source. $m$ is the record's index.
\end{definition}

\begin{definition}[Streaming Graph Record (SGR), $r_m$]\label{def:sgr}
A SR denoted as a quadruple $r_m$=$\langle i,j,\omega,\tau \rangle$, where the payload $p=\langle i,j,\omega\rangle$ indicates an edge with weight $\omega$ between vertices $i$ and $j$.
\end{definition}

The payload can also include an operation such as add (solid edges in Figure~\ref{fig:CDManagementFramework2}) or delete (dashed edges in Figure~\ref{fig:CDManagementFramework2}). For simplicity, we assume all edges are added to the graph and do not consider the operation. 

\begin{definition}[Streaming Graph, $\Re$]\label{def:wbsg}
An unbounded sequence of $SGR$s  denoted as $\Re$=$ \langle r_1, r_2, \cdots \rangle$ in which each record $r_m$ arrives at a destination unit at a particular time $t_m$ ($t_m$$\leq$$t_n$ for $m$$<$$n$).
\end{definition}

\begin{definition}[Burst, $b$]\label{def:burst}
A batch of $S[G]R$s with the same timestamp and the same arrival time. $b$=$\{r_m\mid \nexists r_n: \tau_m$=$\tau_n,t_m$=$t_n, r_n\notin b\}$. 
\end{definition}

We define a burst as the batch of records with the same timestamp which arrive at the computational system \emph{together}. We do not define it as the \emph{all} $SGR$s with the same timestamp, since payloads, timestamps, arrival time points of SGRs, or all of these (i.e., one or more co-arriving SGRs) can be repeated over time due to multiple generative sources and transmission issues. We order the SGRs by both arrival times and generation timestamps. This identifies late arrivals and enables defining a stream as a sequence of arriving bursts, regardless of processing/ingestion window. Figure~\ref{fig:CDManagementFramework2} illustrates a stream with $15$ SGRs arriving at $5$ time-points leading to $7$ bursts:

\small $b_1$=$\{r_1$=$(p_1,\tau_1)$, $r_2$=$(p_2,\tau_1)\}$, $t_1$

\small $b_2$=$\{r_3=(p_3,\tau_2)$,  $r_4$=$(p_4,\tau_2)\}$, $b_3$=$\{r_5$=$(p_1,\tau_1)$, $r_6$=$(p_2,\tau_1)\}$, $t_2$

\small $b_4$=$\{r_7=(p_5,\tau_3)$, $r_8$=$(p_6,\tau_3)\}$, $b_5$=$\{r_9$=$(p_7,\tau_1)$, $r_{10}$=$(p_8,\tau_1)\}$, $t_3$

\small $b_6$=$\{r_{11}$=$(p_9, \tau_4)\}$, $t_4$

\small $b_7$=$\{r_{12}$=$(p_3,\tau_5)$, $r_{13}$=$(p_{10},\tau_5)$, $r_{14}$=$(p_{11},\tau_5)$, $r_{15}$=$(p_{12},\tau_5)\}$, $t_5$

$b_3$ is the same as $b_1$, arriving at a later time-point. Also, $b_5$ has the same timestamp $\tau_1$ as that of $b_1$ and $b_3$, since $r_9$ and $r_{10}$ are late arrivals.  Examples of these cases are when two sources concurrently send the same burst with the same generation time and one ($b_3$, $b_5$) arrives later, or when a networking protocol makes a duplicate ($b_1$) to compensate a delay. 

Note, $r_{15}$ denotes an edge with the same vertices as that of $r_6$,  but with different weight; therefore, the payloads are different; Whereas, $r_{12}\in b_7$ repeats the payload of $r_3\in b_2$, but not the timestamps. Therefore, these are not the same bursts.

\begin{definition}[Window, $W$]\label{def:window}
The set of $SGR$s within an specific interval.
\end{definition}

\begin{definition}[Data Pattern, \th$(W)$]\label{def:sp}
A quantified data characteristic in a window. i.e. \th$(W):W\mapsto \mathbb{R}$.
\end{definition}

\begin{definition}[Transient Concept]\label{def:tc} A non-stable data pattern in data records. i.e. \th$(W)\mid\exists (W_1,t_1),(W_2,t_2):$ \th$(W_1)\neq$ \th$(W_2)$.
\end{definition}

\begin{definition}[Concept Drift (CD)]\label{def:gcd}
The event of a change in a transient concept. 
\end{definition}
Considering a certain pattern \th, concept drift can be detected when observing at least two successive windows $W_1$ and $W_2$ corresponding to sequential time points $t_1$ and $t_2$, where $t_2-t_1\geq1$ and \th$(W_1)\neq$ \th$(W_2)$.

\begin{definition}[Graph Snapshot, $G_{W,t}$]\label{def:graphsnapshot}
The graph $(V,E)$,  formed at time point $t$ by the vertices $V$ and edges $E$ of the $SGR$s within a corresponding window $W$.  
\end{definition}

\begin{definition}[Butterfly, $\bowtie_{j_1,j_2}^{i_1,i_2}$]\label{def:butterfly}
A $(2,2)$-biclique between two i-vertices $i_1$, $i_2$ and two j-vertices $j_1$, $j_2$. It is a closed bipartite four-path $\bowtie_{j_1,j_2}^{i_1,i_2}=\{i_1,j_1, i_2, j_2, i_1 \}$.
\end{definition}

\begin{definition}[Young Butterfly, $\newbowtie$]\label{def:youngbutterfly}
A butterfly with j-vertices having a timestamp within the last $x$ percentage of seen unique timestamps in the stream, i.e. $\newbowtie$=$\{\bowtie_{j_1,j_2}^{i_1,i_2} \mid \exists r_m, r_n: j_1$$\in$$r_m,j_2$$\in$$r_n, (\tau_m,\tau_n)\in[\tau_{t-[xt]},\cdots,\tau_{t-1},\tau_t] \}$.
\end{definition}
Considering young butterflies (i.e. restricting the set of j-vertices), enables case studies where the freshness of input data is important and/or the goal is to perform processing over transient data records rather than all seen data records (streaming processing). This also accounts for the deletions in arriving $SGR$s. We set $x=25\%$. Setting $x=100\%$ would be equivalent to considering all seen vertices.  The set of unique timestamps in the stream grows over time and consequently, the set of j-vertices within the $x$ percentage grows. Choosing a low percentage helps to keep the size of this set balanced particularly when the streaming rate is high.
\section{Literature on CD Detection}\label{sec:relatedworks}
We review the CD management methods through the lens of a modular streaming framework~\cite{lu2019learning} with three components: \underline{data}  \underline{management} for retrieving and retaining data and system state in memory; \underline{drift detection} for identifying changes and corresponding metadata; and \underline{drift adaptation} for updating the downstream task. Accordingly, the existing works are divided in two groups: \textit{active} (Figure~\ref{fig:active}) and \textit{passive} (Figure~\ref{fig:passive}). In active approaches, streaming data is continuously ingested and windowed via data management component and then drifts are explicitly detected and explained via drift detection component. This triggers updating the downstream task via drift adaptation. In passive approaches, a data model is learned in the data management component to extract the most important features of data for dimensionality reduction purposes, and the target goal of the downstream task. Based on the performance of this model (for instance, the learner's error), an implicit drift alert is signaled for drift adaptation. Since our focus is CD detection and understanding, we only review the data management and drift detection components of active and passive approaches. We refer the readers to comprehensive reviews of the works on drift adaptation~\cite{lu2019learning, gama2014survey, agrahari2021concept}. We also do not review the line of works on anomaly detection (e.g.~\cite{Eswaran2018spotlight}). These works identify abnormal data records in known application contexts, while CD signaling is about identifying abnormal situations where a hidden contexts changes and data patterns including concepts change to some extend.

\textbf{Data management.} Data records are continuously ingested and windowed through the window management sub-component and possibly fed into a learner model through the data model sub-component (green boxes in Figures~\ref{fig:active} and \ref{fig:passive}).

\textit{Window management}. While in most passive approaches, a model is learned over a landmark window, active approaches usually use a two-window method with a \textit{reference window} and a \textit{data window}. Contents of data window are evaluated using the reference window as a baseline to determine whether a change has happened. While the data window covers the newly arrived data records, the reference window can be fixed~\cite{shao2014prototype, lu2014concept, bontempelli2022human} or moving~\cite{kifer2004detecting, bach2008paired}. Some active approaches  use single data window. Contents of each window instance are compressed to low dimensional embeddings. This results in a sequence of embeddings as the drift criterion~\cite{yao2016detecting,zambon2018concept,paudel2020approach}. Different techniques have been used for the window borders, window size, and sliding approach. Some approaches use landmark windows~\cite{gama2004learning}, while others use sliding windows~\cite{halstead2022analyzing, zhou2023multi, zhou2024dynamic,nishida2007detecting, yao2016detecting,zambon2018concept,paudel2020approach, bontempelli2022human} with static time-based or count-based sizes~\cite{babcock2002models,yao2016detecting,zambon2018concept,paudel2020approach, bontempelli2022human} or dynamic  sizes~\cite{bifet2007learning, gomes2017adaptive}. When the window size is fixed, all/sampled streaming records are added/removed according to the size~\cite{yao2016detecting,zambon2018concept,paudel2020approach} or a weighting function is used to gradually remove elements with low weights~\cite{gama2014survey}. 

\textit{Data model}. In passive approaches, given a window, a data model is learned which performs the target adaptive task (green boxes in Figure~\ref{fig:passive}). The decrease in model's effectiveness determines the need for an adaptation (yellow box in Figure~\ref{fig:passive}). For instance, when the online error rate of a classifier reaches a drift threshold, a model update is required~\cite{halstead2022analyzing, zhou2023multi, zhou2024dynamic,gama2004learning,liu2017fuzzy,xu2017dynamic,frias2014online}. Some methods also consider a warning threshold to prepare a new model and replace it with the old model when the drift threshold is reached. Some methods involve a human to dismabiguate the drift type before drift adaptation~\cite{bontempelli2022human}. 

\tikzset{arrow/.style = {thick,black,->,>=stealth,}}
\tikzset{arrow2/.style = {thick,dotted,->,>=stealth,}}
\tikzset{nearnodes/.style={node distance=0.5cm}}
\tikzset{farnodes/.style={node distance=2.2cm}}
\tikzset{model1/.style = {rectangle, text width=3cm, minimum height=1cm,text centered, draw=black, fill=green!10,}}
\tikzset{model2/.style = {rectangle, text width=3cm, minimum height=1cm,text centered, draw=black, fill=yellow!25,}}
\tikzset{model4/.style = {rectangle, text width=3cm, minimum height=1cm,text centered, draw=black, fill=blue!25,}}
\tikzset{model3/.style = {}}
\begin{figure}[!ht] 
\centering \resizebox{0.8\linewidth}{!}{
\begin{tikzpicture}

\node[nearnodes] (wm) [model1] {\textbf{Window\\ Management} };

\node[farnodes] (dminput) [model3, left =of wm] {};

\node[farnodes] (eds) [model2, right =of wm] {\textbf{Drift Evaluation}};

\node[nearnodes] (cdc) [model2, above =of eds] {\textbf{Drift Criteria}};

\node[farnodes] (ddoutput) [model3, right =of eds] {};

\node[farnodes] (um) [model4, below =of eds] {\textbf{Model Upgrade} };

\node[farnodes] (daoutput) [model3, right =of um] {};

\node[draw, teal, thick, inner xsep=1em, inner ysep=1em, fit= (wm) ] (box) {};
\node[fill=white] at (box.north) {Data Management};

\node[draw, orange!60, thick, inner xsep=1em, inner ysep=1em, fit=(cdc) (eds)] (box2) {};
\node[fill=white] at (box2.north) {Drift Detection};

\node[draw, blue!60, thick, inner xsep=1em, inner ysep=1em, fit=(um)] (box3) {};
\node[fill=white] at (box3.south) {Drift Adaptation};

\draw [arrow] (cdc.south) -- (eds.north);
\draw [arrow2] (wm.east) -- (cdc.west) node[midway,below]{Samples};
\draw [arrow] (eds.south) -- (um.north) node[midway,right]{Trigger};
\draw [arrow] (eds.east) -- (ddoutput.west) node[midway,above]{ Drift + Descriptions};
\draw [arrow2] (dminput.east) -- (wm.west) node[midway,above]{Stream};
\draw [arrow2] (um.east) -- (daoutput.west) node[midway,above]{Output};
\draw [arrow] (um.west) -- (box.south east);
\end{tikzpicture}}
\caption{Active concept drift management.} \label{fig:active}
\end{figure}
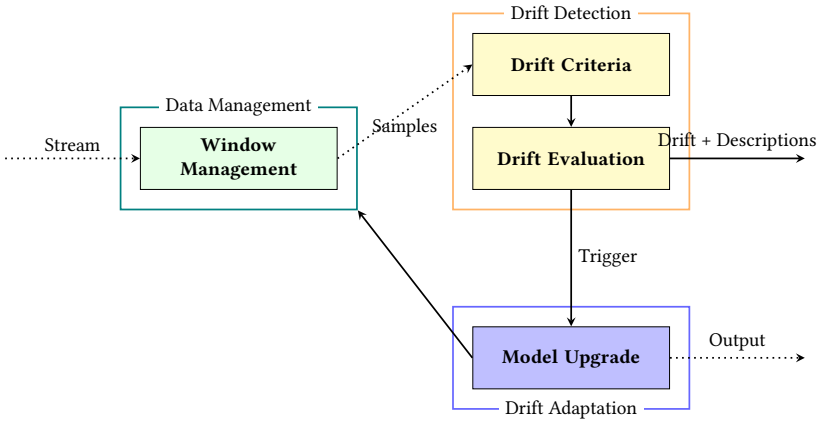
\tikzset{arrow/.style = {thick,black,->,>=stealth,}}
\tikzset{arrow2/.style = {thick,dotted,->,>=stealth,}}
\tikzset{nearnodes/.style={node distance=0.7cm}}
\tikzset{farnodes/.style={node distance=2.2cm}}
\tikzset{model1/.style = {rectangle, text width=3cm, minimum height=1cm,text centered, draw=black, fill=green!10,}}
\tikzset{model2/.style = {rectangle, text width=3cm, minimum height=1cm,text centered, draw=black, fill=yellow!25,}}
\tikzset{model4/.style = {rectangle, text width=3cm, minimum height=1cm,text centered, draw=black, fill=blue!25,}}
\tikzset{model3/.style = {}}
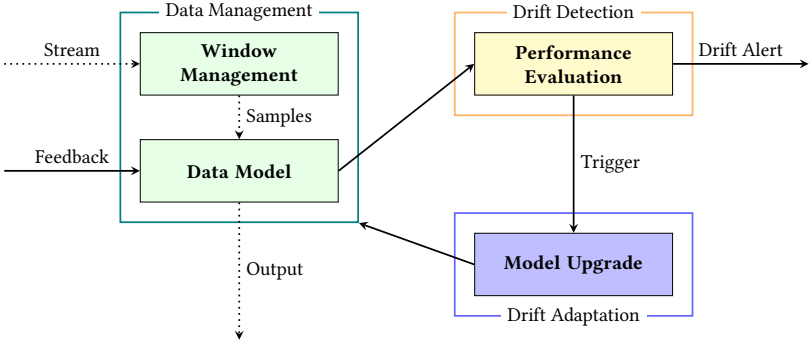
\begin{figure} [!ht]
\centering
\resizebox{0.8\linewidth}{!}{
\begin{tikzpicture}

\node[nearnodes] (wm) [model1] {\textbf{Window\\ Management} };
\node[farnodes] (dminput) [model3, left =of wm] {};

\node[nearnodes] (dm) [model1, below =of wm] {\textbf{Data Model} };
\node[farnodes] (feedback) [model3, left =of dm] {};
\node[farnodes] (dmoutput) [model3, below =of dm] {};

\node[farnodes] (ep) [model2, right =of wm] {\textbf{Performance Evaluation}};

\node[farnodes] (ddoutput) [model3, right =of ep] {};

\node[farnodes] (um) [model4, below =of ep] {\textbf{Model Upgrade} };

\node[draw, teal, thick, inner xsep=1em, inner ysep=1em, fit= (wm)(dm) ] (box) {};
\node[fill=white] at (box.north) {Data Management};

\node[draw, orange!60, thick, inner xsep=1em, inner ysep=1em, fit=(ep)] (box2) {};
\node[fill=white] at (box2.north) {Drift Detection};

\node[draw, blue!60, thick, inner xsep=1em, inner ysep=1em, fit=(um)] (box3) {};
\node[fill=white] at (box3.south) {Drift Adaptation};

\draw [arrow2] (wm.south) -- (dm.north) node[midway,right]{Samples};
\draw [arrow] (dm.east) -- (ep.west) ;
\draw [arrow] (ep.south) -- (um.north) node[midway,right]{Trigger};
\draw [arrow] (ep.east) -- (ddoutput.west) node[midway,above]{Drift Alert};
\draw [arrow] (feedback.east) -- (dm.west) node[midway,above]{Feedback};
\draw [arrow2] (dminput.east) -- (wm.west) node[midway,above]{Stream};
\draw [arrow2] (dm.south) -- (dmoutput.north) node[midway,right]{Output};
\draw [arrow] (um.west) -- (box.south east) ;
\end{tikzpicture}}
\caption{Passive concept drift management.} \label{fig:passive}
\end{figure}

\textbf{Drift detection.} 
In active approaches, CD is usually detected when a statistical property of the data stream changes over time. The first formal definition of change detection in data streams~\cite{kifer2004detecting} considers windows as data samples and computes their distribution distance to identify a drift using a hypothesis test method. This type of drift detection is also done using multiple hypothesis tests running in parallel or as a hierarchy of sequential tests~\cite{zhang2017three, yu2018request}. A recent line of research on graph streams (sequence of attributed or labeled graph snapshots) convert the graph stream to a time-series and perform change detection over it. The elements of time-series are prototype-based graph embedding vectors~\cite{zambon2018concept}, or entropy of discriminative subgraphs with respect to classification labels~\cite{yao2016detecting} or with respect to a minimum description length~\cite{paudel2020approach}. The drift is evaluated by measuring a diversion dissimilarity~\cite{paudel2020approach} or a hypothesis test~\cite{zambon2018concept} and utilising a static threshold~\cite{zambon2018concept, paudel2020approach, yao2016detecting}. 

As we explain in the next sections, \emph{SGDP} employs an active approach that extracts effective knowledge from the SGRs (the burstiness of the SGRs) on-the-fly, and efficiently maintain it as the system state. \emph{SGDD} combines active and passive approaches and summarizes the stream into a graph of butterflies which is further reduced to two time-series (similarity of the butterfly neighborhoods and similarity of the future changes in butterfly neighborhoods) as the system state. Both \emph{SGDP} and \emph{SGDD} use a single data window and do not use reference window. The data window is a burst-based landmark window sliding with the arrival of each burst. In each window instance, the ingested timestamp of the newly arrived burst and the updated average burst size are captured in respective data collections. The SGR payloads are windowed in \emph{SGDD} but not in \emph{SGDP} and out-of-order timestamps are not captured repeatedly. \emph{SGDP} and \emph{SGDD} examine the time-series for CD signals. They both use dynamic thresholds set according to the number of detections and the streaming rate status.




\section{SGDP}\label{sec:sGradP}
\emph{SGDP} analyzes the time-series of burst sizes to signal an upcoming CD. This reduces the problem of CD signaling in streaming graphs to  the problem of change detection in the time-series of burst sizes.  

The functional architecture of \emph{SGDP} is similar to that of active approaches (Figure~\ref{fig:active}). The  main framework is given in Algorithm~\ref{alg:SGDP}.
\emph{SGDP} performs two main tasks: 

    \paragraph{\textcolor{teal}{Data Management}} This component extracts the burstiness properties of the stream and uses a sliding window with an adaptive length set to one burst, to append the current burst size to a time-series and regulate the frequency of analyses. 
    \emph{SGDP} just reads the generation timestamp of SGRs to extract and update the burstiness profile of the stream. It can even use a hash map of these timestamps. Hence, with minimum access to the data records and a light-weight time-series change detection, it predicts the change in the generative source of streaming (graph) data. 
    
    \paragraph{\textcolor{orange!60}{Drift Detection}}  The second component analyzes a suffix of the time series to check for upcoming CDs. Previous studies have shown that the characteristic substructure (butterflies) in streaming graphs emerge through bursty addition of edges to the graph~\cite{sheshbolouki2022sgrapp, Sheshbolouki:2023aa}. Therefore, \emph{SGDP} detects a change in the burstiness patterns as an indication of abnormal generation processes. The frequency of these change detections depends on the streaming rate since the analyses start at the arrival of each burst (i.e., each window instance with a length adapting to the streaming rate).
\begin{algorithm}[!ht]\caption{SGDP()}\label{alg:SGDP}
  \DontPrintSemicolon
    \KwData{$\Re= \langle r_1, r_2, \cdots \rangle$,  sequence of sgrs}
    $uniqueTimestamps \gets \emptyset$,
    $W \gets 1$, $t\gets 1$, $B\gets 1$, $\bar{B}\gets 0$, $B_{max}\gets 0$, $Bcount\gets 0$, $\bar{B}series\gets \emptyset$, $W_{d}series\gets \{0\}$ \\          %
    
    \While{$\exists r_t=\langle p_m, \tau_m \rangle$}{

        $Bcount\gets uniqueTimestamps.size()$\label{alg:SGDP:burstprofilestart}
       
       \If{$uniqueTimestamps \ni \tau_t$}{
            $B++$
        }
        \Else{
            $\bar{B}\gets (\bar{B}\times Bcount+B)/(|Bcount|+1)$\\
            $B\gets 1$
        }
        \If{$B>B_{max}$}{
            $B_{max}\gets B$\\
            System.gc()\label{alg:SGDP:gc}
        }\label{alg:SGDP:burstprofileend}
        
        \If{($\tau_t\notin uniqueTimestamps$ $\&$ $Bcount>1$)}{\label{alg:SGDP:taskstart}
            $\bar{B}series$.add($\bar{B}$)\\\label{alg:SGDP:bseries}
            $uniqueTimestamps.add(\tau_t)$\\\label{alg:SGDP:timestampadd}
            
            \ForAll{ $f \in \langle 1, 0.1, 0.9, 0.2, 0.8, 0.3, 0.7, 0.4, 0.6, 0.5\rangle$}{\label{alg:SGDP:f}
               \If{$W-W_{d}series.lastElement()>\bar{B}$}{
                    $CDS_{Bursts}$($B_{max},\bar{B},\bar{B}series,W,t,W_{d}series,f$)\\
                }
            }  
            $W++$\label{alg:SGDP:taskend}
        }
        \Else{$uniqueTimestamps.add(\tau_t)$\\}
        $t++$
    }
\end{algorithm}
\subsection{SGDP - Data Management}\label{subsec:datamanagement_SGDP}

The window management is done as follows.

    \textbf{\textcolor{teal}{1)}} When a SGR arrives, the burstiness profile of the stream is updated on the fly (Algorithm~\ref{alg:SGDP}, Lines~\ref{alg:SGDP:burstprofilestart}-\ref{alg:SGDP:burstprofileend}). This includes the number of seen bursts ($Bcount$), the size of current burst ($B$), the average burst size ($\Bar{B}$), and the maximum seen burst size ($B_{max}$).

    \textbf{\textcolor{teal}{2)}} A new timestamp denotes a new burst (a new window instance with a dynamic size), which initiates CD check analysis over $\Bar{B}series$ (Algorithm~\ref{alg:SGDP}, Lines~\ref{alg:SGDP:taskstart}-\ref{alg:SGDP:taskend}). The streaming rate of data records could be highly dynamic. While existing works disregard the generation timestamps of edges and analyze graph snapshots with a fix window size, \emph{SGDP} uses burst-based windows with a size adapting to the burst sizes (streaming rate). We use this adaptive window length to resolve the following scalability problems of time-based windows with fixed sizes. When the streaming rate is high, either the window drops data records through sampling or sliding (trading the accuracy), or the window is split to sub-windows~\cite{zhang24incremental} and each sub-window is processed independently (losing the inter-connections among data or performing extra processing). E.g., consider a high degree vertex in a large graph with skewed degree distribution; where the neighbours of the vertex fall in disjoint sub-windows. Solving this issue requires further graph partitioning processes or double checking the connections between sub-windows, which defeats the efficiency purpose.  When the rate is low, the window should wait for the arrival of data records to start the analyses (trading throughput). 
    
    When a new timestamp arrives, the updated average burst size is appended to a time series ($\Bar{B}series$) and the timestamp is added to a hash set of unique values (Algorithm~\ref{alg:SGDP}, Lines~\ref{alg:SGDP:bseries}-\ref{alg:SGDP:timestampadd}). If a repeated timestamp arrives (a late arrival), it would not be captured again. 

    \textbf{\textcolor{teal}{3)}} Next, if the last CD signal has been in at least a distance of $\Bar{B}$ previous windows, $\Bar{B}series$ is analyzed for a CD signal at the current window (i.e., current burst). This analysis requires a threshold factor $f$. In our experiments we realized that most CD signals are issued when $f=0.3$, therefore we just use this value. However other values can be tried for $f$ (as suggested in  Algorithm~\ref{alg:SGDP}, Line~\ref{alg:SGDP:f}).

\subsection{SGDP - Drift Detection}\label{subsec:driftdetection_SGDP}

CD check is done as follows:

    \textbf{\textcolor{orange!60}{1)}} To check for a CD signal, the last $S$ elements of $\Bar{B}serie$ are examined. This suffix size $S$ increases as a function of the maximum and average burst sizes (Algorithm~\ref{alg:checkdriftburst}, Line~\ref{Suffixsize}).

    \textbf{\textcolor{orange!60}{2)}} If there are enough elements in  $\Bar{B}serie$, then the number of elements within the suffix that are greater and less than the current average burst size ($Ngreater$, $Nless$) are captured  (Algorithm~\ref{alg:checkdriftburst}, Lines~\ref{nstart}-\ref{nend}).

    \textbf{\textcolor{orange!60}{3)}} If any of these values passes a dynamic threshold ($\ceil{S\times f}$), a CD is signaled and the current window (burst) number is added to its corresponding time series $W_dseries$ (Algorithm~\ref{alg:checkdriftburst}, Lines~\ref{signalstart}-\ref{signalend}).
\begin{algorithm}[!ht]\caption{CD check via Burst sizes}\label{alg:checkdriftburst}
  \DontPrintSemicolon
   \Fn{\hypertarget{CDSBursts}{$CDC_{Bursts}$($B_{max},\bar{B},\bar{B}series,W,t,W_{d}series,f$)}}{
        $d\gets W_{d}series.size+1$\\
        $S \gets maxB\frac{\floor{log_{10}(Max(maxB,100))}}{\floor{log_{10}(Max(\Bar{B},10)}})$\\ \label{Suffixsize}
        $s2\gets\bar{B}series.size()$, 
        $Ngreater \gets 0$, 
        $Nless \gets0$

        \If{$s2>S$}{
            \ForAll{$i\gets s2-S-1$; $i<s2$; $i++$}{\label{nstart}
                $temp\gets \bar{B}series.elementAt(i)$\\
                \If{$temp>\bar{B}$}{$Ngreater++$}
                \If{$temp<\bar{B}$}{$Nless++$}
            }\label{nend}
            $threshold\gets \ceil{S\times f}$\\
            \If{$Ngreater\geq threshold \bigvee Nless \geq threshold$}{
               $W_{d}series.add(W)$\\ \label{signalstart}
               Signal a drift at the sgr index $t$, window (burst) number $W$, and current system time\label{signalend}
            }
        }
    }
\end{algorithm}
\section{Performance Evaluation}\label{sec:evaluations}

\paragraph{Data.} We simulate streaming graphs with ground truth about drift's time and pattern in the experiments. We synthesize $10^6$ SGRs by the \emph{sGrow} model~\cite{Sheshbolouki:2023aa}, with a prefix of $1000$ real-world SGRs from Amazon user-item stream\footnote{Available at public repositories KONECT \url{konect.uni-koblenz.de/networks/} and Netzschleuder \url{networks.skewed.de}}. \emph{sGrow} as a configurable model generates bursts  from several concurrent origins such that the streaming graph reproduces realistic subgraph emergence patterns. We simulate a change in a hidden context (change in the generative process) rather than a change in a target concept (e.g., subgraph inter-connectivity patterns). We refer to the switch from the $1000$th real-world SGR to the first synthetic SGR as the first 
CD and simulate the next 
CDs as the following. 

We introduce changes to the \emph{sGrow}'s generative process via changing two parameters of \emph{sGrow} which contribute the most to the emergence of butterflies: $[L_{min}, L_{max}]$ (range of preferential random walk's dynamic lengths), and $\rho$ (burst connection probability). Two other parameters of the model (window parameter $\beta$ and batch size $M$) are fixed. Parameters are set as the following:
\begin{itemize}
    \item $M$ and $\beta$ can be set to any user-specified value without affecting the characteristic patterns of generated stream. we use the default values $\beta=5$ and $M=10$ in the experiments.
    \item  The default value for $\rho$ is $0.3$ and for $[L_{min},L_{max}]$ is $[1,2]$. Increasing $\rho$ to values less than $0.7$ and expanding the range of $[L_{min},L_{max}]$ ensure preserving butterfly emergence patterns, while decreasing the generation time and increasing burst size. We use $\rho=0.4$ and $[L_{min},L_{max}]=[1,4]$ as the initial values to reduce the generation time while preserving realistic patterns and leaving room for drift simulation.
\end{itemize}

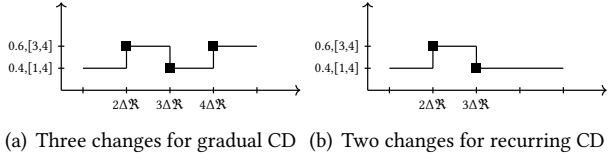
\begin{figure}[!h]
    \centering
\subfigure[Three changes for gradual CD]{\resizebox{4cm}{!}{
\begin{tikzpicture}[thick,bnode/.style={fill=black}]
\draw[thick,->] (0,0) -- node[below=3ex] {} (5.5,0) ;
\draw[thick,->] (0,0) -- node[left=3em,rotate=90]  {} (0,2) ;
\draw (0.5,2pt) -- ++ (0,-4pt) node[below, draw=none] {};
\draw (1.5,2pt) -- ++ (0,-4pt) node[below] {$2\Delta\Re$};
\draw (2.5,2pt) -- ++ (0,-4pt) node[below, draw=none] {$3\Delta\Re$};
\draw (3.5,2pt) -- ++ (0,-4pt) node[below] {$4\Delta\Re$};
\draw (4.5,2pt) -- ++ (0,-4pt) node[below, draw=none] {};

\draw (2pt,0.5) -- ++ (-4pt,0) node[left]  {0.4,[1,4]};
\draw (2pt,1.0) -- ++ (-4pt,0) node[left]  {0.6,[3,4]};

\node[bnode,xshift=15mm,yshift=10mm] ()[] {};

\node[bnode,xshift=25mm,yshift=5mm] () [] {};

\node[bnode,xshift=35mm,yshift=10mm] () [] {};

\draw[thick,-] (0.5,0.5) -- node[below=3ex] {} (1.5,0.5) ;
\draw[thick,-] (1.5,0.5) -- node[below=3ex] {} (1.5,1) ;
\draw[thick,-] (1.5,1) -- node[below=3ex] {} (2.5,1) ;
\draw[thick,-] (2.5,1) -- node[below=3ex] {} (2.5,0.5) ;
\draw[thick,-] (2.5,0.5) -- node[below=3ex] {} (3.5,0.5) ;
\draw[thick,-] (3.5,0.5) -- node[below=3ex] {} (3.5,1) ;
\draw[thick,-] (3.5,1) -- node[below=3ex] {} (4.5,1) ;

\end{tikzpicture}}}
\subfigure[Two changes for recurring CD]{\resizebox{4cm}{!}{
\begin{tikzpicture}[thick,bnode/.style={fill=black}]
\draw[thick,->] (0,0) -- node[below=3ex] {} (5.5,0) ;
\draw[thick,->] (0,0) -- node[left=3em,rotate=90]  {} (0,2) ;
\draw (0.5,2pt) -- ++ (0,-4pt) node[below, draw=none] {};
\draw (1.5,2pt) -- ++ (0,-4pt) node[below] {$2\Delta\Re$};
\draw (2.5,2pt) -- ++ (0,-4pt) node[below, draw=none] {$3\Delta\Re$};
\draw (3.5,2pt) -- ++ (0,-4pt) node[below] {};
\draw (4.5,2pt) -- ++ (0,-4pt) node[below, draw=none] {};
\draw (2pt,0.5) -- ++ (-4pt,0) node[left]  {0.4,[1,4]};
\draw (2pt,1.0) -- ++ (-4pt,0) node[left]  {0.6,[3,4]};

\draw[thick,-] (0.5,0.5) -- node[below=3ex] {} (1.5,0.5) ;
\draw[thick,-] (1.5,0.5) -- node[below=3ex] {} (1.5,1) ;
\draw[thick,-] (1.5,1) -- node[below=3ex] {} (2.5,1) ;
\draw[thick,-] (2.5,1) -- node[below=3ex] {} (2.5,0.5) ;
\draw[thick,-] (2.5,0.5) -- node[below=3ex] {} (4.5,0.5) ;


\node[bnode,xshift=15mm, yshift=10mm] (a2) [] {};

\node[bnode,xshift=25mm, yshift=5mm] (a3) [] {};

\end{tikzpicture}}}
    \caption{Evolution of \emph{sGrow}'s parameters over the timeline of SGR generation. $\Delta\Re =10^5, 2\times 10^5$ is the drift interval in terms of the number of generated SGRs.}
    \label{fig:driftPatterns}
\end{figure}
We simulate gradual and recurring drift patterns by switching the parameters of \emph{sGrow} according to Figure~\ref{fig:driftPatterns}. For gradual CD, the transient concept changes gradually and frequently, while spanning a considerable time interval until a new concept is stabilized. For recurring CD, the transient concept switches to a new concept and then it is repeated. We use drift intervals of $\Delta\Re= 10^5$ and $2\times10^5$ SGRs. We record the timestamp at which the drift is introduced for the evaluations. 

Five stream instances are generated per pattern per drift interval for a total of $20$ streams. We denote the streams as $R_{ab}$ and $G_{ab}$, where 
\begin{itemize}
    \item $R$ refers to recurring drifts.
    \item $G$ refers to gradual drifts.
    \item $a=1$ refers to $\Delta\Re= 1\times10^5$ (close-drift stream).
    \item $a=2$ refers to $\Delta\Re= 2\times 10^5$ (far-drift stream).
    \item $b\in\{1,2,3,4,5\}$ refers to the stream's instance number.
\end{itemize}
The length of the stream suffix without CD varies from $400K$ SGRs ($R_{2b}$ and $G_{2b}$), to $600K$ SGRs ($G_{1b}$), and $700K$ SGRs ($R_{1b}$).

\paragraph{Metrics.}
CD signals are issued discretely. For each CD, we calculate the average system time distance (ms) and the SGR count distance between the CD and the first and last signals before that CD. The SGR count distance is fixed over multiple execution of the algorithms over a data stream, however the system time varies. We calculate the average time distances with $100$ executions over each data stream since the standard deviation of the execution times is stabilized after $100$ executions. We run the algorithm in $10$ separate batch of $10$ executions to overcome the caching/operating system effects on the performance.

\paragraph{Computing setup.} We conduct the experiments with $15.6$ GB native memory and Intel Core $i7 - 6770HQ CPU @ 2.60GHz * 8$ processor. All algorithms are implemented in Java (OpenJDK $17.0.12$).

\paragraph{Baseline.} Existing works on CD detection do not operate on streaming graphs, therefore we introduce a baseline framework, called \emph{SGDD}, for streaming graph concept drift detection.
\emph{SGDD} represents the streaming graph as an evolving network of butterflies and tracks the similarity of neighbourhood of butterflies to signal CD. This reduces the problem of CD detection in bipartite streaming graphs to change detection in time series of similarity values.

\emph{SGDD}'s functional architecture is shown in Figure~\ref{fig:proposed} and the main framework is given in Algorithm \ref{alg:framework}. \emph{SGDD} performs two main tasks combining the architectures of active and passive approaches. Detailed descriptions are provided in Appendices~\ref{sGraddDataManagement} and \ref{sGraddDriftDetection}. 
\tikzset{arrow/.style = {thick,black,->,>=stealth}}
\tikzset{arrow2/.style = {thick,dotted,->,>=stealth}}
\tikzset{nearnodes/.style={node distance=1.5cm}}
\tikzset{farnodes/.style={node distance=2.4cm}}
\tikzset{invisiblenodes/.style={node distance=1.5cm}}
\tikzset{model1/.style = {rectangle, text width=5cm, minimum height=1cm,text centered, draw=black, fill=green!10}}
\tikzset{model2/.style = {rectangle, text width=5cm, minimum height=1cm,text centered, draw=black, fill=yellow!25}}
\tikzset{model3/.style = {}}
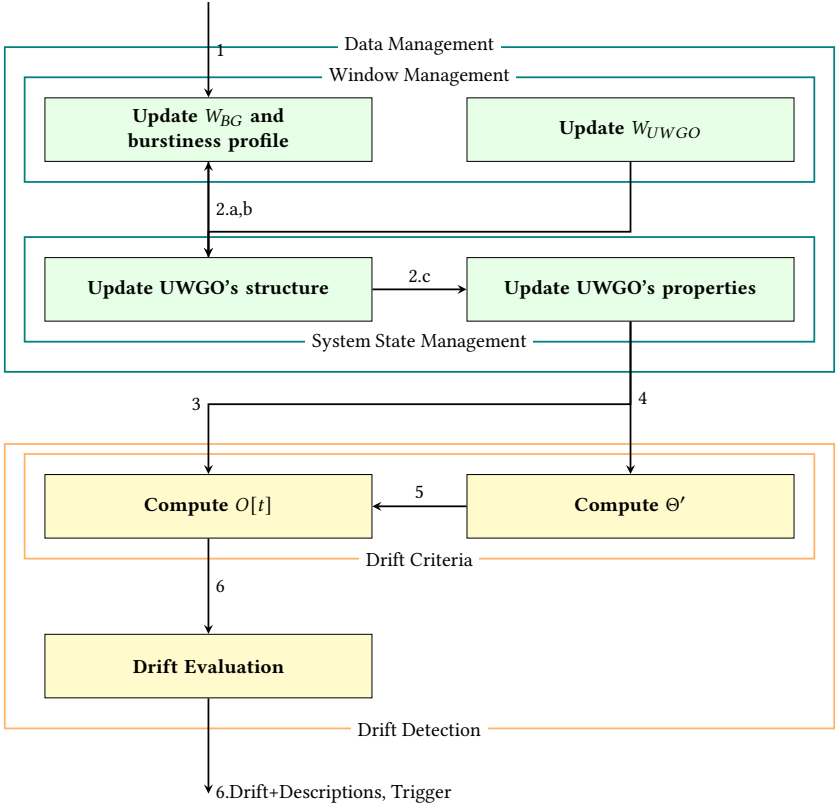
\begin{figure}[!ht]
\centering
\resizebox{0.8\linewidth}{!}{
\begin{tikzpicture}

\node[nearnodes] (wm) [model1] {\textbf{Update $W_{BG}$ and burstiness profile} };
\node[nearnodes] (wm2) [model1, right =of wm] {\textbf{Update $W_{UWGO}$} };

\node[nearnodes] (dm) [model1, below =of wm] {\textbf{Update UWGO's structure} };
\node[nearnodes] (dm2) [model1, right =of dm] {\textbf{Update UWGO's properties} };

\node[invisiblenodes] (dminput) [model3, above =of wm] {};

\node[farnodes] (cdc) [model2, below =of dm] {\textbf{Compute $O[t]$}};
\node[nearnodes] (cdc2) [model2, right =of cdc] {\textbf{Compute $\Theta^\prime$}};

\node[nearnodes] (eds) [model2, below =of cdc] {\textbf{Drift Evaluation}};

\node[invisiblenodes] (ddoutput) [model3, right =of eds] {};

\node[invisiblenodes] (trigger) [model3, below =of eds] {};

\node[invisiblenodes] (daoutput) [model3, right =of um] {};

\node[draw, teal, thick, inner xsep=2em, inner ysep=2.5em, fit= (wm)(wm2)(dm)(dm2) ] (box) {};
\node[fill=white] at (box.north) {Data Management};

\node[draw, teal, thick, inner xsep=1em, inner ysep=1em, fit= (wm)(wm2) ] (box11) {};
\node[fill=white] at (box11.north) {Window Management};

\node[draw, teal, thick, inner xsep=1em, inner ysep=1em, fit= (dm)(dm2) ] (box12) {};
\node[fill=white] at (box12.south) {System State Management};

\node[draw, orange!60, thick, inner xsep=2em, inner ysep=1.5em, fit=(cdc)(cdc2) (eds)] (box2) {};
\node[fill=white] at (box2.south) {Drift Detection};

\node[draw, orange!60, thick, inner xsep=1em, inner ysep=1em, fit=(cdc)(cdc2)] (box21) {};
\node[fill=white] at (box21.south) {Drift Criteria};

\draw [arrow] (dminput.south) -- (wm.north) node[midway,right]{1};
\draw [arrow] (wm.south) -- (dm.north) node[midway,right]{2.a,b};
\draw [arrow] (dm.east) -- (dm2.west) node[midway,above]{2.c};
\draw [arrow] (dm.north) -- (wm.south) node[midway,right]{};
\draw [arrow] (wm2.south) to++(0,-1.1)-| (dm.north) node[right=7,above]{};
\draw [arrow] (dm2.south) -- (cdc2.north) node[midway,right]{4};
\draw [arrow] (dm2.south) to++(0,-1.3)-| (cdc.north) node[midway,left]{3};
\draw [arrow] (cdc2.west) -- (cdc.east) node[midway,above]{5};
\draw [arrow] (cdc.south) -- (eds.north) node[midway,right]{6};
\draw [arrow] (eds.south) -- (trigger.north) node[right]{6.Drift+Descriptions, Trigger};

\end{tikzpicture}}
\caption{\emph{SGDD}'s architecture.} \label{fig:proposed}
\end{figure}
\paragraph{\textcolor{teal}{Data Management}} 
The goal is efficiently extracting and maintaining the state of the transient concept in the streaming graph (butterfly interconnectivity patterns).  To this end, SGRs are ingested from the \textbf{b}ipartite streaming \textbf{g}raph to a burst-based sliding window $W_{BG}$ and the burstiness profile of the stream is updated on the fly. At the arrival of each burst, $W_{BG}$ is projected to a predicate-based sliding window $W_{UWGO}$ which contains a \textbf{u}nipartite \textbf{w}eighted \textbf{g}raph of \textbf{o}scillators. Each vertex in UWGO is an oscillator and represents a young butterfly in $W_{BG}$ with an oscillating phase denoting the butterfly's dynamic neighbourhood. The edge weights denote the neighbourhood sizes at the time of establishing connections among incident butterflies. The intuition is that the neighbourhood of a butterfly fluctuates between zero to $N-1$ neighbours (where $N$ is the number of butterflies in UWGO). Therefore, a butterfly's neighbourhood is represented by an oscillating phase with a frequency of oscillation. Same butterfly neighbourhoods are hashed to the same phase. Therefore, $UWGO$ summarizes the interconnectivity of butterfly motifs in the original streaming graph (transient concept). 
$W_{UWGO}$ has low computational overhead since it is updated incrementally. Moreover, $W_{BG}$ is entirely retired as soon as it is projected to $W_{UWGO}$ (i.e. it is a tumbling window). This (a) frees up memory for the drift detection since deleted objects are collected by the garbage collector and (b) avoids redundancy since in the next instance of $W_{BG}$, there wouldn't be any existing butterfly to be enumerated again. Also, $W_{BG}$ and $W_{UWGO}$ adapt to the streaming rate by adjusting the slide size to the burst sizes.
\paragraph{\textcolor{orange!60}{ Drift Detection} }
A drift is detected by detecting a change in the drift criteria. The degree of global synchronization of phases in UWGO (similarity of butterfly neighbourhoods) reflects the density of butterfly interconnections (the relative size of the largest complete subgraph), and a change in the emergence of butterfly motifs indicates a change in the generative source(s) of the stream. Therefore, we analyze the synchronization of phases as a drift criterion. The challenge is that when the generative source changes, it takes a while for butterflies to form and connect according to the new generative condition and consequently the transient concept (UWGO) displays one or scattered changes with a delay. Therefore, just relying on the currently observed state of the transient concept and reporting a CD at the observation of a change in the synchronization would create false, duplicate, and delayed CD detections, while we want to signal each drift  as soon as possible. To address this, we predict phase changes and analyze the synchronization of both the phases and their changes.
\begin{table*}[!ht]
    \caption{The average system time distance (ms) \textcolor{cyan}{/} \textcolor{c5}{SGR count distance} of the ith CD and its first and last SGDD's signals.  
}\centering\resizebox{\textwidth}{!}{
    \begin{tabular}{|c|c|c|c|c|c|c|c|c|c|c|}
        \hline \tiny ms\textcolor{cyan}{/}\textcolor{c5}{SGR}&\cellcolor{yellow}$d_{1f}$&\cellcolor{yellow}$d_{1l}$  &   \cellcolor{orange}$d_{2f}$&\cellcolor{orange}$d_{2l}$    &\cellcolor{mustard}$d_{3f}$&\cellcolor{mustard}$d_{3l}$    &\cellcolor{darkorange}$d_{4f}$&\cellcolor{darkorange}$d_{4l}$ &  \cellcolor{cherrybrown}$d_{5f}$&\cellcolor{cherrybrown}$d_{5l}$ \\\hline\hline
        \cellcolor{green1}$G_{11}$   &$1279.22\textcolor{cyan}{/}\textcolor{c5}{982}$&$1279.22\textcolor{cyan}{/}\textcolor{c5}{982}$& &&  $875.45\textcolor{cyan}{/}\textcolor{c5}{19167}$&$33.99\textcolor{cyan}{/}\textcolor{c5}{988}$& $8537.4\textcolor{cyan}{/}\textcolor{c5}{98732}$&$124.79\textcolor{cyan}{/}\textcolor{c5}{1864}$& $51.4\textcolor{cyan}{/}\textcolor{c5}{876}$&$22731.01\textcolor{cyan}{/}\textcolor{c5}{594829}$ \\
        \cellcolor{green1}$G_{12}$   &$1286.93\textcolor{cyan}{/}\textcolor{c5}{982}$&$1286.93\textcolor{cyan}{/}\textcolor{c5}{982}$& &&  && && $224.74\textcolor{cyan}{/}\textcolor{c5}{2468}$&$29664.04\textcolor{cyan}{/}\textcolor{c5}{597297}$ \\
        \cellcolor{green1}$G_{13}$   &$1318.93\textcolor{cyan}{/}\textcolor{c5}{982}$&$1318.93\textcolor{cyan}{/}\textcolor{c5}{982}$& &&  && && $19796.47\textcolor{cyan}{/}\textcolor{c5}{208221}$&$43807.36\textcolor{cyan}{/}\textcolor{c5}{596554}$ \\
        \cellcolor{green1}$G_{14}$   &$1305.08\textcolor{cyan}{/}\textcolor{c5}{982}$&$1305.08\textcolor{cyan}{/}\textcolor{c5}{982}$& &&  && && $7467.8\textcolor{cyan}{/}\textcolor{c5}{87587}$&$42014.77\textcolor{cyan}{/}\textcolor{c5}{600536}$ \\
        \cellcolor{green1}$G_{15}$   &$1316.48\textcolor{cyan}{/}\textcolor{c5}{982}$&$1316.48\textcolor{cyan}{/}\textcolor{c5}{982}$& &&  $503.58\textcolor{cyan}{/}\textcolor{c5}{10751}$&$0.35\textcolor{cyan}{/}\textcolor{c5}{233}$& $8920.9\textcolor{cyan}{/}\textcolor{c5}{98988}$&$115.65\textcolor{cyan}{/}\textcolor{c5}{1300}$& $61.09\textcolor{cyan}{/}\textcolor{c5}{570}$&$23105.18\textcolor{cyan}{/}\textcolor{c5}{594899}$ \\
        \hline\hline
        \cellcolor{green1}$G_{21}$   &$1345.31\textcolor{cyan}{/}\textcolor{c5}{982}$&$1345.31\textcolor{cyan}{/}\textcolor{c5}{982}$& &&  $910.22\textcolor{cyan}{/}\textcolor{c5}{16152}$&$188.32\textcolor{cyan}{/}\textcolor{c5}{4558}$& $25925.09\textcolor{cyan}{/}\textcolor{c5}{198732}$&$89.56\textcolor{cyan}{/}\textcolor{c5}{638}$& $153.99\textcolor{cyan}{/}\textcolor{c5}{1843}$&$25642.7\textcolor{cyan}{/}\textcolor{c5}{397538}$ \\
        \cellcolor{green1}$G_{22}$   &$1311.41\textcolor{cyan}{/}\textcolor{c5}{982}$&$1311.41\textcolor{cyan}{/}\textcolor{c5}{982}$& &&  $4581.48\textcolor{cyan}{/}\textcolor{c5}{84374}$&$65.07\textcolor{cyan}{/}\textcolor{c5}{921}$& $22908.16\textcolor{cyan}{/}\textcolor{c5}{198384}$&$64.05\textcolor{cyan}{/}\textcolor{c5}{505}$& $172.88\textcolor{cyan}{/}\textcolor{c5}{2310}$&$22775.03\textcolor{cyan}{/}\textcolor{c5}{399502}$ \\
        \cellcolor{green1}$G_{23}$   &$1381.85\textcolor{cyan}{/}\textcolor{c5}{982}$&$1381.85\textcolor{cyan}{/}\textcolor{c5}{982}$& &&  $5327.45\textcolor{cyan}{/}\textcolor{c5}{134239}$&$66.48\textcolor{cyan}{/}\textcolor{c5}{1825}$& $16039.6\textcolor{cyan}{/}\textcolor{c5}{198208}$&$0.9\textcolor{cyan}{/}\textcolor{c5}{721}$& $207.83\textcolor{cyan}{/}\textcolor{c5}{4561}$&$15697.04/395782$ \\
        \cellcolor{green1}$G_{24}$   &$1324.19\textcolor{cyan}{/}\textcolor{c5}{982}$&$1324.19\textcolor{cyan}{/}\textcolor{c5}{982}$& &&  $4879.08\textcolor{cyan}{/}\textcolor{c5}{105844}$&$108.54\textcolor{cyan}{/}\textcolor{c5}{2335}$& $20195.31\textcolor{cyan}{/}\textcolor{c5}{199897}$&$182.62\textcolor{cyan}{/}\textcolor{c5}{2067}$& $49.68\textcolor{cyan}{/}\textcolor{c5}{1236}$&$20383.99\textcolor{cyan}{/}\textcolor{c5}{395417}$ \\
        \cellcolor{green1}$G_{25}$   &$1338.6\textcolor{cyan}{/}\textcolor{c5}{982}$&$1338.6\textcolor{cyan}{/}\textcolor{c5}{982}$& &&  $4648.39\textcolor{cyan}{/}\textcolor{c5}{113738}$&$25.84\textcolor{cyan}{/}\textcolor{c5}{2125}$& $17670.39\textcolor{cyan}{/}\textcolor{c5}{197973}$&$75.8\textcolor{cyan}{/}\textcolor{c5}{956}$& $131.42\textcolor{cyan}{/}\textcolor{c5}{3098}$&$17632.63\textcolor{cyan}{/}\textcolor{c5}{397629}$ \\\hline
        
        \tiny AVG& 1320.8\textcolor{cyan}{/}\textcolor{c5}{982} &1320.8\textcolor{cyan}{/}\textcolor{c5}{982}  \textcolor{cyan}{/}\textcolor{c5}{} && &3103.7\textcolor{cyan}{/}\textcolor{c5}{69180.7} &69.8\textcolor{cyan}{/}\textcolor{c5}{1855}& 17171\textcolor{cyan}{/}\textcolor{c5}{170132}& 69.8\textcolor{cyan}{/}\textcolor{c5}{1150.1}&
             2831.7\textcolor{cyan}{/}\textcolor{c5}{31277}    & 26345.4\textcolor{cyan}{/}\textcolor{c5}{496998.3} \\
         \hline
        \end{tabular}
        }
        \resizebox{\textwidth}{!}{
        \begin{tabular}{|c|c|c|c|c|c|c|c|c|}
        \hline \tiny ms\textcolor{cyan}{/}\textcolor{c5}{SGR}&\cellcolor{yellow}$d_{1f}$&\cellcolor{yellow}$d_{1l}$  &   \cellcolor{orange}$d_{2f}$&\cellcolor{orange}$d_{2l}$    &\cellcolor{mustard}$d_{3f}$&\cellcolor{mustard}$d_{3l}$   &\cellcolor{cherrybrown}$d_{4f}$&\cellcolor{cherrybrown}$d_{4l}$ \\\hline\hline
        \cellcolor{lightpurple}$R_{11}$   &$1475.07\textcolor{cyan}{/}\textcolor{c5}{982}$&$1475.07\textcolor{cyan}{/}\textcolor{c5}{982}$&  && && $5935.04\textcolor{cyan}{/}\textcolor{c5}{57671}$&$75446.28\textcolor{cyan}{/}\textcolor{c5}{600966}$ \\
        \cellcolor{lightpurple}$R_{12}$   &$1447.28\textcolor{cyan}{/}\textcolor{c5}{982}$&$1447.28\textcolor{cyan}{/}\textcolor{c5}{982}$&  && && $19451.59\textcolor{cyan}{/}\textcolor{c5}{158393}$&$92599.45\textcolor{cyan}{/}\textcolor{c5}{664706}$ \\
        \cellcolor{lightpurple}$R_{13}$   &$1539.74\textcolor{cyan}{/}\textcolor{c5}{982}$&$1539.74\textcolor{cyan}{/}\textcolor{c5}{982}$&  && && $8017.77\textcolor{cyan}{/}\textcolor{c5}{71784}$&$80846.53\textcolor{cyan}{/}\textcolor{c5}{602937}$ \\
        \cellcolor{lightpurple}$R_{14}$   &$1332.92\textcolor{cyan}{/}\textcolor{c5}{982}$&$1332.92\textcolor{cyan}{/}\textcolor{c5}{982}$&  && $$&$$& $17077.71\textcolor{cyan}{/}\textcolor{c5}{146204}$&$89826.2\textcolor{cyan}{/}\textcolor{c5}{686197}$ \\
        \cellcolor{lightpurple}$R_{15}$   &$1418.04\textcolor{cyan}{/}\textcolor{c5}{982}$&$1418.04\textcolor{cyan}{/}\textcolor{c5}{982}$& &&  $2849.95\textcolor{cyan}{/}\textcolor{c5}{68446}$&$64.32\textcolor{cyan}{/}\textcolor{c5}{2566}$& $105.79\textcolor{cyan}{/}\textcolor{c5}{1768}$&$58976.76\textcolor{cyan}{/}\textcolor{c5}{596884}$ \\
        \hline\hline
        \cellcolor{lightpurple}$R_{21}$   &$1381.17\textcolor{cyan}{/}\textcolor{c5}{982}$&$1381.17\textcolor{cyan}{/}\textcolor{c5}{982}$& $11046.09\textcolor{cyan}{/}\textcolor{c5}{98705}$&$86.94\textcolor{cyan}{/}\textcolor{c5}{1267}$&  $9405.91\textcolor{cyan}{/}\textcolor{c5}{199209}$&$106.17\textcolor{cyan}{/}\textcolor{c5}{3318}$& $114.57\textcolor{cyan}{/}\textcolor{c5}{915}$&$40008.31\textcolor{cyan}{/}\textcolor{c5}{398823}$\\
        \cellcolor{lightpurple}$R_{22}$   &$1376.46\textcolor{cyan}{/}\textcolor{c5}{982}$&$1376.46\textcolor{cyan}{/}\textcolor{c5}{982}$& &&  $4329.28\textcolor{cyan}{/}\textcolor{c5}{66057}$&$151.64\textcolor{cyan}{/}\textcolor{c5}{4311}$& $128.32\textcolor{cyan}{/}\textcolor{c5}{893}$&$44998.35\textcolor{cyan}{/}\textcolor{c5}{398598}$\\
        \cellcolor{lightpurple}$R_{23}$   &$1354.48\textcolor{cyan}{/}\textcolor{c5}{982}$&$1354.48\textcolor{cyan}{/}\textcolor{c5}{982}$& &&  $6531.81\textcolor{cyan}{/}\textcolor{c5}{93877}$&$223.41\textcolor{cyan}{/}\textcolor{c5}{4712}$& $47.33\textcolor{cyan}{/}\textcolor{c5}{286}$&$51448.63\textcolor{cyan}{/}\textcolor{c5}{398240}$\\
        \cellcolor{lightpurple}$R_{24}$   &$1349.71\textcolor{cyan}{/}\textcolor{c5}{982}$&$1349.71\textcolor{cyan}{/}\textcolor{c5}{982}$&  $$&$$& $$&$$& $35030.2\textcolor{cyan}{/}\textcolor{c5}{222298}$&$68710.73\textcolor{cyan}{/}\textcolor{c5}{399689}$ \\
        \cellcolor{lightpurple}$R_{25}$   &$1348.72\textcolor{cyan}{/}\textcolor{c5}{982}$&$1348.72\textcolor{cyan}{/}\textcolor{c5}{982}$& &&  $8990.3\textcolor{cyan}{/}\textcolor{c5}{103329}$&$127.73\textcolor{cyan}{/}\textcolor{c5}{2970}$& $124.7\textcolor{cyan}{/}\textcolor{c5}{675}$&$55020.2\textcolor{cyan}{/}\textcolor{c5}{398164}$ \\
        \hline
        \tiny AVG& 1402.3\textcolor{cyan}{/}\textcolor{c5}{982} & 1402.3\textcolor{cyan}{/}\textcolor{c5}{982}  & 11046.1\textcolor{cyan}{/}\textcolor{c5}{98705}& 86.9\textcolor{cyan}{/}\textcolor{c5}{1267}&    6421.4\textcolor{cyan}{/}\textcolor{c5}{106183.6}& 134.6\textcolor{cyan}{/}\textcolor{c5}{3575.4}& 8603.3
             \textcolor{cyan}{/}\textcolor{c5}{66088.7}& 65788.1\textcolor{cyan}{/}\textcolor{c5}{514520.4}\\
         \hline
    \end{tabular}}
    
    \label{tab:SGDD_dist}
\end{table*}
\begin{table*}[!ht]
    \caption{The average system time distance (ms) \textcolor{cyan}{/} \textcolor{c5}{SGR count distance} of the ith CD and its first and last \emph{SGDP}'s signals  : $d_{if}$, $d_{il}$ in sGrow streams with \textcolor{green1}{gradual} and \textcolor{lightpurple}{recurring} CD. The last two columns refer to the signals after the last CD.}\centering\resizebox{\textwidth}{!}{
    \begin{tabular}{|c|c|c|c|c|c|c|c|c|c|c|}
        \hline  \tiny ms\textcolor{cyan}{/}\textcolor{c5}{SGR}&\cellcolor{yellow}$d_{1f}$&\cellcolor{yellow}$d_{1l}$  &   \cellcolor{orange}$d_{2f}$&\cellcolor{orange}$d_{2l}$    &\cellcolor{mustard}$d_{3f}$&\cellcolor{mustard}$d_{3l}$    &\cellcolor{darkorange}$d_{4f}$&\cellcolor{darkorange}$d_{4l}$ &  \cellcolor{cherrybrown}$d_{5f}$&\cellcolor{cherrybrown}$d_{5l}$ \\\hline\hline
							
         \cellcolor{green1}$G_{11}$&    651.42\textcolor{cyan}{/}\textcolor{c5}{990}&	584.01\textcolor{cyan}{/}\textcolor{c5}{801}&		2762.67\textcolor{cyan}{/}\textcolor{c5}{197322}&	111.69\textcolor{cyan}{/}\textcolor{c5}{496}&		305.15\textcolor{cyan}{/}\textcolor{c5}{79689}&	8.51\textcolor{cyan}{/}\textcolor{c5}{23465}& && &\\

         \cellcolor{green1}$G_{12}$&   683.68\textcolor{cyan}{/}\textcolor{c5}{990}&607.8\textcolor{cyan}{/}\textcolor{c5}{801}&		3006.75\textcolor{cyan}{/}\textcolor{c5}{198233}&1.32\textcolor{cyan}{/}\textcolor{c5}{3349}&		2463.45\textcolor{cyan}{/}\textcolor{c5}{91772}&1933.86\textcolor{cyan}{/}\textcolor{c5}{25163}&		26.12\textcolor{cyan}{/}\textcolor{c5}{63150}&5.67\textcolor{cyan}{/}\textcolor{c5}{13125}&&\\ 
         	
         \cellcolor{green1}$G_{13}$&   609.44\textcolor{cyan}{/}\textcolor{c5}{990}&535.87\textcolor{cyan}{/}\textcolor{c5}{801}&		1607.15\textcolor{cyan}{/}\textcolor{c5}{198294}&	0.96\textcolor{cyan}{/}\textcolor{c5}{2086}&		3150.85\textcolor{cyan}{/}\textcolor{c5}{95263}&3.58\textcolor{cyan}{/}\textcolor{c5}{10180}&		39.02\textcolor{cyan}{/}\textcolor{c5}{98054}&0.19\textcolor{cyan}{/}\textcolor{c5}{292}&		7.2\textcolor{cyan}{/}\textcolor{c5}{17716}&1029.1\textcolor{cyan}{/}\textcolor{c5}{189478}\\
         
         \cellcolor{green1}$G_{14}$&   598.39\textcolor{cyan}{/}\textcolor{c5}{990}&534.11\textcolor{cyan}{/}\textcolor{c5}{801}&		2170.12\textcolor{cyan}{/}\textcolor{c5}{198157}&1.99\textcolor{cyan}{/}\textcolor{c5}{5219}&		1758.48\textcolor{cyan}{/}\textcolor{c5}{98704}&544.96\textcolor{cyan}{/}\textcolor{c5}{12354}&		33.05\textcolor{cyan}{/}\textcolor{c5}{95479}&2.8\textcolor{cyan}{/}\textcolor{c5}{7674}&		2\textcolor{cyan}{/}\textcolor{c5}{5404}&25.15\textcolor{cyan}{/}\textcolor{c5}{70548}\\
         
         \cellcolor{green1}$G_{15}$&    648.8\textcolor{cyan}{/}\textcolor{c5}{990}&575.54\textcolor{cyan}{/}\textcolor{c5}{801}&		3405\textcolor{cyan}{/}\textcolor{c5}{198337}&	1.13\textcolor{cyan}{/}\textcolor{c5}{2978}&		2272.32\textcolor{cyan}{/}\textcolor{c5}{86842}&1056.86\textcolor{cyan}{/}\textcolor{c5}{58633}&    &&  &\\
         
         \hline\hline
        
         \cellcolor{green1}$G_{21}$& 566.15\textcolor{cyan}{/}\textcolor{c5}{990}&484.24\textcolor{cyan}{/}\textcolor{c5}{801}&		2391.25\textcolor{cyan}{/}\textcolor{c5}{198283}&1.08\textcolor{cyan}{/}\textcolor{c5}{2278}&		4687.6\textcolor{cyan}{/}\textcolor{c5}{193215}&1389.33\textcolor{cyan}{/}\textcolor{c5}{32314}& && &\\
         
         \cellcolor{green1}$G_{22}$&   641.91\textcolor{cyan}{/}\textcolor{c5}{990}&	572.85\textcolor{cyan}{/}\textcolor{c5}{801}&		2687.29\textcolor{cyan}{/}\textcolor{c5}{198289}&	2.46\textcolor{cyan}{/}\textcolor{c5}{7023}&		4017.14\textcolor{cyan}{/}\textcolor{c5}{196268}&	2293.06\textcolor{cyan}{/}\textcolor{c5}{111395}& && &\\
         
         \cellcolor{green1}$G_{23}$&    631.01\textcolor{cyan}{/}\textcolor{c5}{990}&	553.63\textcolor{cyan}{/}\textcolor{c5}{801}&		3268.29\textcolor{cyan}{/}\textcolor{c5}{198489}&	1.12\textcolor{cyan}{/}\textcolor{c5}{3069}&		5695.47\textcolor{cyan}{/}\textcolor{c5}{184920}&	5695.47\textcolor{cyan}{/}\textcolor{c5}{184920}& && &\\
         
         \cellcolor{green1}$G_{24}$&    656.74\textcolor{cyan}{/}\textcolor{c5}{990}&	586.69\textcolor{cyan}{/}\textcolor{c5}{801}&		3078.45\textcolor{cyan}{/}\textcolor{c5}{198172}&	1.11\textcolor{cyan}{/}\textcolor{c5}{3151}&		3758.52\textcolor{cyan}{/}\textcolor{c5}{190072}&	2739.9\textcolor{cyan}{/}\textcolor{c5}{147150}&   && &\\
         		
         \cellcolor{green1}$G_{25}$&   717.63\textcolor{cyan}{/}\textcolor{c5}{990}&	639.03\textcolor{cyan}{/}\textcolor{c5}{801}&		3300.2\textcolor{cyan}{/}\textcolor{c5}{198171}&	1.64\textcolor{cyan}{/}\textcolor{c5}{4135}&		1683.72\textcolor{cyan}{/}\textcolor{c5}{187192}&	1675.52\textcolor{cyan}{/}\textcolor{c5}{164620}&    &&  &\\
         \hline
         \tiny AVG& 640.52\textcolor{cyan}{/}\textcolor{c5}{990}&567.38\textcolor{cyan}{/}\textcolor{c5}{801}  &2767.72\textcolor{cyan}{/}\textcolor{c5}{198174.7}&12.45\textcolor{cyan}{/}\textcolor{c5}{3378.4}&    2979.27\textcolor{cyan}{/}\textcolor{c5}{140393.7}&1734.1\textcolor{cyan}{/}\textcolor{c5}{77019.4}&
             32.73\textcolor{cyan}{/}\textcolor{c5}{85561}&2.89\textcolor{cyan}{/}\textcolor{c5}{7030.33}&  4.6\textcolor{cyan}{/}\textcolor{c5}{11560}&527.12\textcolor{cyan}{/}\textcolor{c5}{130013} \\
         \hline
    \end{tabular}
   }
    \centering\resizebox{\textwidth}{!}{
    \begin{tabular}{|c|c|c|c|c|c|c|c|c|}
        \hline \tiny ms\textcolor{cyan}{/}\textcolor{c5}{SGR}& \cellcolor{yellow}$d_{1f}$&\cellcolor{yellow}$d_{1l}$  &   \cellcolor{orange}$d_{2f}$&\cellcolor{orange}$d_{2l}$    &\cellcolor{mustard}$d_{3f}$&\cellcolor{mustard}$d_{3l}$    &\cellcolor{darkorange}$d_{4f}$&\cellcolor{darkorange}$d_{4l}$ \\\hline\hline
					
        \cellcolor{lightpurple}$R_{11}$&    636.49\textcolor{cyan}{/}\textcolor{c5}{990}&558.96\textcolor{cyan}{/}\textcolor{c5}{801}&		2462.21\textcolor{cyan}{/}\textcolor{c5}{198127}&2.45\textcolor{cyan}{/}\textcolor{c5}{6869}&		4278.58\textcolor{cyan}{/}\textcolor{c5}{97673}&4278.58\textcolor{cyan}{/}\textcolor{c5}{97673}&		250.91\textcolor{cyan}{/}\textcolor{c5}{682266}&250.91\textcolor{cyan}{/}\textcolor{c5}{682266}\\
         									
        \cellcolor{lightpurple}$R_{12}$&    561.16\textcolor{cyan}{/}\textcolor{c5}{990}&485.26\textcolor{cyan}{/}\textcolor{c5}{801}&		2409.21\textcolor{cyan}{/}\textcolor{c5}{198394}&1.47\textcolor{cyan}{/}\textcolor{c5}{3558}&		976.68\textcolor{cyan}{/}\textcolor{c5}{94770}&	1.93\textcolor{cyan}{/}\textcolor{c5}{5188}&		3.11\textcolor{cyan}{/}\textcolor{c5}{9031}&268.26\textcolor{cyan}{/}\textcolor{c5}{697540}\\
        
        \cellcolor{lightpurple}$R_{13}$&    542.94\textcolor{cyan}{/}\textcolor{c5}{990}&472.12\textcolor{cyan}{/}\textcolor{c5}{801}&		2604.77\textcolor{cyan}{/}\textcolor{c5}{198448}&0.6\textcolor{cyan}{/}\textcolor{c5}{1219}&		2217.9\textcolor{cyan}{/}\textcolor{c5}{86849}&1084.84\textcolor{cyan}{/}\textcolor{c5}{31365}&		219.16\textcolor{cyan}{/}\textcolor{c5}{651394}&	233.98\textcolor{cyan}{/}\textcolor{c5}{691495}\\
        
        \cellcolor{lightpurple}$R_{14}$&    549.44\textcolor{cyan}{/}\textcolor{c5}{990}&478.82\textcolor{cyan}{/}\textcolor{c5}{801}&		2796.4\textcolor{cyan}{/}\textcolor{c5}{198409}&2.29\textcolor{cyan}{/}\textcolor{c5}{6153}&		3512.81\textcolor{cyan}{/}\textcolor{c5}{99174}&1854.34\textcolor{cyan}{/}\textcolor{c5}{20925}&		21.43\textcolor{cyan}{/}\textcolor{c5}{63266}&	251.95\textcolor{cyan}{/}\textcolor{c5}{691375}\\
        
        \cellcolor{lightpurple}$R_{15}$&    637.72\textcolor{cyan}{/}\textcolor{c5}{990}&560.35\textcolor{cyan}{/}\textcolor{c5}{801}&		3779.05\textcolor{cyan}{/}\textcolor{c5}{198169}&3.08\textcolor{cyan}{/}\textcolor{c5}{8142}&		4835.85\textcolor{cyan}{/}\textcolor{c5}{93607}&	4835.85\textcolor{cyan}{/}\textcolor{c5}{93607}&&\\
								
         \hline\hline
         
         \cellcolor{lightpurple}$R_{21}$&   699.96\textcolor{cyan}{/}\textcolor{c5}{990}&615.46\textcolor{cyan}{/}\textcolor{c5}{801}&		5225.88\textcolor{cyan}{/}\textcolor{c5}{398443}&5.35\textcolor{cyan}{/}\textcolor{c5}{13644}&		7374.02\textcolor{cyan}{/}\textcolor{c5}{195310}&	7374.02\textcolor{cyan}{/}\textcolor{c5}{195310}&&\\
		
         \cellcolor{lightpurple}$R_{22}$&   630.29\textcolor{cyan}{/}\textcolor{c5}{990}&558.23\textcolor{cyan}{/}\textcolor{c5}{801}&		3453.2\textcolor{cyan}{/}\textcolor{c5}{398254}&2.2\textcolor{cyan}{/}\textcolor{c5}{4616}&		6982.72\textcolor{cyan}{/}\textcolor{c5}{192559}&	32.63\textcolor{cyan}{/}\textcolor{c5}{84623}&&\\
         
         \cellcolor{lightpurple}$R_{23}$&  543.12\textcolor{cyan}{/}\textcolor{c5}{990}&475.34\textcolor{cyan}{/}\textcolor{c5}{801}&		3502.31\textcolor{cyan}{/}\textcolor{c5}{398270}&2.92\textcolor{cyan}{/}\textcolor{c5}{8113}&		3924.48\textcolor{cyan}{/}\textcolor{c5}{196196}&1690.2\textcolor{cyan}{/}\textcolor{c5}{122332}&		130.03\textcolor{cyan}{/}\textcolor{c5}{384054}&130.03\textcolor{cyan}{/}\textcolor{c5}{384054}\\
				
         \cellcolor{lightpurple}$R_{24}$&  544.31\textcolor{cyan}{/}\textcolor{c5}{990}&474.32\textcolor{cyan}{/}\textcolor{c5}{801}&		2520.09\textcolor{cyan}{/}\textcolor{c5}{398463}&1.87\textcolor{cyan}{/}\textcolor{c5}{4604}&		3696.13\textcolor{cyan}{/}\textcolor{c5}{196815}&0.89\textcolor{cyan}{/}\textcolor{c5}{2159}&		4.31\textcolor{cyan}{/}\textcolor{c5}{11958}&166.67\textcolor{cyan}{/}\textcolor{c5}{398053}\\
	         
         \cellcolor{lightpurple}$R_{25}$&  539.37\textcolor{cyan}{/}\textcolor{c5}{990}&471.82\textcolor{cyan}{/}\textcolor{c5}{801}&		3712.9\textcolor{cyan}{/}\textcolor{c5}{398096}&1.1\textcolor{cyan}{/}\textcolor{c5}{2813}&		3373.39\textcolor{cyan}{/}\textcolor{c5}{186715}&1212.24\textcolor{cyan}{/}\textcolor{c5}{114096}&		95.07\textcolor{cyan}{/}\textcolor{c5}{280736}&135.51\textcolor{cyan}{/}\textcolor{c5}{383225} \\
         \hline
         \tiny AVG& 588.48\textcolor{cyan}{/}\textcolor{c5}{990}&515.07\textcolor{cyan}{/}\textcolor{c5}{801}&3246.6\textcolor{cyan}{/}\textcolor{c5}{298307.3}&2.33\textcolor{cyan}{/}\textcolor{c5}{5973.1}&4117.26\textcolor{cyan}{/}\textcolor{c5}{143966.8}&2236.55\textcolor{cyan}{/}\textcolor{c5}{76727.8}&103.43\textcolor{cyan}{/}\textcolor{c5}{297529.14}&205.33\textcolor{cyan}{/}\textcolor{c5}{561144}\\
         \hline
    \end{tabular}
    }
   \label{tab:sgradp_dist}
\end{table*}

\paragraph{Results.}
When the generative source of a streaming graph changes, \emph{SGDD} signals CD with a notable delay especially when the change points are closer ($R_{1b}$ and $G_{1b}$ rows in Table~\ref{tab:SGDD_dist}). This delay is due to the time required for listing and analyzing butterflies and makes it difficult to distinguish the false positive signals (missed signals) and false negative signals (delayed signals). On the other hand, as shown in Table~\ref{tab:sgradp_dist}, \emph{SGDP} discerns the synthetic SGRs from real-world SGRs (first CD) by issuing signals within SGR distance range of $990-810$ and average system time distance of $640.52-567.38$/$588.48-515.07$ (ms). Regarding the parameter changes, CD2 is approximately predicted between $3$ seconds to $2$ milliseconds  before it occurs ($198,000$ or $398,0000$ to variant numbers of SGRs). CD3 signals are issued from approximately $4$ seconds until $2$ seconds before it occurs (with different SGR distances across streams). And CD4 for $G_{1b}$ streams is signaled from approximately $32$ milliseconds until $2.89$ milliseconds before its occurrence. False positive signals are more frequent for $R_{ab}$ streams since the parameter drop leads to higher burst sizes making the CD checks analyzing a larger number of burst sizes due to the increase of maximum burst size and $S$, while the threshold increases. False negatives happen in $G_{ab}$ streams only.  We evaluated the performance of \emph{SGDP} with the suffix size determined as $\frac{\floor{log_{10}(Max(maxB,100))}}{\floor{log_{10}(Max(\Bar{B},10)}})^{(-1)^{d}}$ (Table~\ref{tab:SGDP0}). We observed this removal of the Maximum burst size from calculation for $S$, increases the false positives and also signals span almost the entire timeline of the SGR arrivals.
\section{Conclusion}\label{sec:conclusion}
Concept drift is a natural phenomenon in streaming graphs. We define transient concepts and drift in streaming graphs. Our definitions enable studying diverse data patterns and concepts. We take the butterfly inter-connectivity patterns and introduce a framework for streaming graph drift detection, called \emph{SGDD} which signals the drifts at the observation of butterfly neighbourhoods' tendency to change. \emph{SGDD}'s data management techniques, which include identifying butterflies and the relative size of the largest complete network of butterflies, display promising visions for identifying maximal subgraphs and summarizing streaming graphs. We also focus on the burstiness characteristics of the stream and introduce an advanced framework for streaming graph drift \emph{prediction}, called \emph{SGDP}, which combines on-the-fly data management with minimum access to data records and light-weight drift prediction techniques. Both \emph{SGDD} and \emph{SGDP} can be integrated with any online adaptive analytics; they are designed as unsupervised techniques for understanding and detecting drifts in hidden contexts (generative sources) which are reflected in data patterns. While \emph{SGDD} detects CDs with a delay, \emph{SGDP} discerns the synthetic SGRs from real-world SGRs by issuing predictive signals within a  distance range of $990-810$ SGRs and average system time distance of $640.52-567.38$ (ms) or $588.48-515.07$ (ms). \emph{SGDP} predicts parameter changes of the generative process(s) within a system time distance range of $\approx4 (s)$ to $\approx2.89(ms)$ (starting at $\approx198,000$ or $398,0000$ SGRs before the CD occurrence). \emph{SGDP} can achieve false positives of zero in streams with gradual CD pattern. False negatives happen in streams with gradual CD patterns, but not in the streams with recurring CD pattern.  Our experiments show this false positive/negative rates can be improved with further tuning the signaling algorithm.

\bibliographystyle{ACM-Reference-Format}
\bibliography{CDreferences,publications, references-sGrow}

\appendix

\section{Appendices}\label{appendixA}

\subsection{SGDD - Data Management}\label{sGraddDataManagement}
The interleaved procedures of window management and system state management are as follows (steps 1 and 2 and green boxes in Figure~\ref{fig:proposed}).
\begin{algorithm}[!ht]\caption{SGDD()}\label{alg:framework}
  \DontPrintSemicolon
    \KwData{$\Re= \langle r_1, r_2, \cdots \rangle$,  sequence of sgrs}
    
    $BBG \gets \emptyset$, 
    $UWGO = (V, E) \gets \emptyset$, 
    $uniqueTimestamps \gets \emptyset$,
    $W \gets 1$, $t\gets 1$, $B\gets 1$, $\bar{B}\gets 0$, $maxB\gets 0$, $Bcount\gets1$,  $d\gets 0$, $W_dserie\gets\{0\}$
    
    \While{$\exists r_t=\langle i_t, j_t, \omega_t, \tau_t \rangle)$}{
       $Bcount\gets uniqueTimestamps.size()$\label{alg:cd:burstprofilestart}
       
       \If{$uniqueTimestamps \ni \tau_t$}{
            $B++$
        }
        \Else{
            $\bar{B}\gets (\bar{B}\times Bcount+B)/(|Bcount|+1)$\\
            $B\gets 1$
        }
        \If{$B>B_{max}$}{
            $B_{max}\gets B$\\
            System.gc()\label{alg:cd:gc1}
        }\label{alg:cd:burstprofileend}

        \If{$BG\notni (i_t,j_t)$}{\label{alg:ingeststart}
        $BG.addedge(i_t,j_t)$\\}\label{alg:ingestend}
        \If{($uniqueTimestamps \notni \tau_t$ $\&$  $Bcount>1$)}{\label{alg:cd:taskstart}
            $uniqueTimestamps.add(\tau^t)$
            
            Project $BG$ to $UWGO$\label{alg:cd:project}
            
            $BG \gets \emptyset$ \label{alg:cd:BGrenew}
            
            \ForAll{ $v\in V$}{\label{alg:cd:assignphasefreqstart}
                $\theta_v \gets (\Sigma_{n\in N(v)} n.ID) \% 2\pi$ 
                
                $\Omega_v\gets$ sample from a Gaussian distribution\label{alg:cd:assignphasefreqend}                
            }  
            
            System.gc()\label{alg:cd:gc2}
           
            $O_1[W]\gets ((\sum_{v\in V} sin \theta_v)^2+(\sum_{v\in V} cos \theta_v)^2)^{\frac{1}{2}}/|V|$\label{mainalg:o1}
            
            $\Delta\theta \gets  RungeKutta(UWGO,0.01)$\label{mainalg:theta2}
            
            $O_2[W]\gets ((\sum_{v\in V} sin \Delta\theta_v)^2+(\sum_{v\in V} cos \Delta\theta_v)^2)^{\frac{1}{2}}/|V|$\label{mainalg:o2}
            
            $CDC_{Butterfly}(\bar{B},maxB,O_1,O_2,t,W, W_dserie)$\label{mainalg:driftdetect}
           
            $W++$\label{alg:cd:retireUWGOend}
        }
        \Else{$uniqueTimestamps.add(\tau^t)$}
        
        $t++$
    }
\end{algorithm}

\begin{algorithm}[!ht]\caption{Project $W_{BG}$ to $W_{UWGO}$}\label{alg:project}
  \DontPrintSemicolon
  \ForAll{$\bowtie_v=\{i_1,j_1, i_2, j_2, i_1 \} \in BG$}{
        $v\gets$ new UWGO vertex with $\theta_v =0$, $\Omega_v =0$, $ID=0$\label{alg:project:mapstart}\\
        $v.setID(v.hashCode())$\\
        $UWGO.add(v)$\label{alg:project:mapend}\\
        Add young butterflies adjacent to $j_1$ and $j_2$ to L\label{alg:project:addedgestart}\\ 
        \ForAll{$\bowtie_u\in L$}{
                $UWGO.addEdge( u,v,|L| )$\label{alg:project:addedgeend}
        }
    }
    Local data structures $\gets \emptyset$\label{alg:project:setNull}\\
    System.gc()\label{alg:project:gc}\\    
   \end{algorithm}

    \textbf{\textcolor{teal}{1)}} Arriving SGRs are added to $W_{BG}$ and the burstiness profile of the stream 
    is updated online (Algorithm~\ref{alg:framework}, lines~\ref{alg:cd:burstprofilestart}-\ref{alg:ingestend}
    ). When one burst is seen, $W_{BG}$ is closed and the following steps 2 to 7 are performed (Algorithm~\ref{alg:framework}, lines~\ref{alg:cd:taskstart}-\ref{alg:cd:retireUWGOend}). 
    
    \textbf{\textcolor{teal}{2)}} $W_{BG}$ is projected to update $W_{UWGO}$.

    \textbf{\textcolor{teal}{2.a)}} The  UWGO structure is updated by identifying the young butterflies, mapping them to UWGO vertices, and connecting these vertices (Algorithm\ref{alg:framework}, line~\ref{alg:cd:project} invoking Algorithm~\ref{alg:project}). The young butterflies are identified using the exact algorithm in the \emph{sGrapp} suit~\cite{sheshbolouki2022sgrapp}. Figure~\ref{fig:projectBGtoUWGO}(a) illustrates 8 young butterflies in an instance of $W_{BG}$ (butterflies incident to $j_0$ are excluded):  
    \begin{equation*}\bowtie_{v_1} = \bowtie_{j_1,j_2}^{i_2,i_3}, \bowtie_{v_2} = \bowtie_{j_1,j_2}^{i_2,i_4}, \bowtie_{v_3} = \bowtie_{j_1,j_2}^{i_3,i_4}, \bowtie_{v_4} = \bowtie_{j_2,j_3}^{i_5,i_6},\bowtie_{v_5} = \bowtie_{j_4,j_5}^{i_7,i_8}, \bowtie_{v_6} = \bowtie_{j_5,j_6}^{i_9,i_{10}} \end{equation*} 
    \begin{equation*}\bowtie_{v_7} = \bowtie_{j_6,j_7}^{i_{11},i_{12}}, \bowtie_{v_8} = \bowtie_{j_8,j_9}^{i_{13},i_{14}} \end{equation*}

\begin{figure}[!ht]\centering

\subfigure[BG snapshot. Young butterflies (solid lines) are connected through j-vertices.]{\resizebox{\linewidth}{!}{
    \begin{tikzpicture}[thick,
  every node/.style={draw,circle},
  bnode/.style={fill=blue},
  ynode/.style={fill=yellow},
  every fit/.style={ellipse,draw,inner sep=-2pt,text width=2cm}]
\begin{scope}[xshift=0.9cm,start chain=going right,node distance=10mm]
    \node[on chain] (f0) [label=below: $j_{0}$] {};
    \node[bnode,on chain] (f1) [label=below: $j_1$] {};
    \node[bnode,on chain] (f2)[label=above: $j_2$] {};
    \node[bnode,on chain] (f3) [label=above: $j_3$] {};
\end{scope}
\begin{scope}[xshift=5.5cm,start chain=going right,node distance=10mm]
    \node[bnode,on chain] (f4) [label=below: $j_4$] {};
    \node[bnode,on chain] (f5) [label=below: $j_5$] {};
    \node[bnode,on chain] (f6) [label=right: $j_6$] {};
     \node[bnode,on chain] (f7) [label=above: $j_7$] {};
      \node[bnode,on chain] (f8) [label=below: $j_8$] {};
      \node[bnode,on chain] (f9) [label=below: $j_9$] {};
      
\end{scope}

\begin{scope}[xshift=1.2cm,yshift=10mm,start chain=going right,node distance=5mm]
    \node[ynode,on chain] (s2) [label=above: $i_2$] {};
    \node[ynode,on chain] (s3) [label=above: $i_3$] {};
    \node[ynode,on chain] (s4) [label=above: $i_4$] {};
\end{scope}

\begin{scope}[xshift=3.5cm,yshift=-10mm,start chain=going right,node distance=5mm]
    \node[ynode,on chain] (s5) [label=below: $i_5$] {};
    \node[ynode,on chain] (s6) [label=below: $i_6$] {};
\end{scope}
\begin{scope}[xshift=5.5cm,yshift=10mm,start chain=going right,node distance=5mm]
    \node[ynode,on chain] (s7) [label=above: $i_7$] {};
    \node[ynode,on chain] (s8) [label=above: $i_8$] {};
\end{scope}
\begin{scope}[xshift=8
cm,yshift=10mm,start chain=going right,node distance=5mm]
    \node[ynode,on chain] (s9) [label=above: $i_9$] {};
    \node[ynode,on chain] (s10) [label=above: $i_{10}$] {};
\end{scope}
\begin{scope}[xshift=8.5cm,yshift=-10mm,start chain=going right,node distance=5mm]
    \node[ynode,on chain] (s11) [label=below: $i_{11}$] {};
    \node[ynode,on chain] (s12) [label=below: $i_{12}$] {};    
\end{scope}
\begin{scope}[xshift=11.5cm,yshift=10mm,start chain=going right,node distance=5mm]
    \node[ynode,on chain] (s13) [label=above: $i_{13}$] {};
    \node[ynode,on chain] (s14) [label=above: $i_{14}$] {};
\end{scope}

\draw[dotted] (f7) -- (s13);
\draw[dotted] (f3) -- (s7);
\draw[dotted] (f0) -- (s2);
\draw[dotted]  (f0) -- (s3);

\draw  (f1) -- (s2);
\draw  (f2) -- (s2);
\draw  (f1) -- (s3);
\draw  (f2) -- (s3);
\draw  (f1) -- (s4);
\draw  (f2) -- (s4);
\draw  (f2) -- (s5);
\draw  (f3) -- (s5);
\draw  (f2) -- (s6);
\draw  (f3) -- (s6);

\draw  (f4) -- (s7);
\draw  (f5) -- (s8);
\draw  (f4) -- (s8);
\draw  (f5) -- (s7);
\draw  (f5) -- (s9);
\draw  (f5) -- (s10);
\draw  (f6) -- (s10);
\draw  (f6) -- (s9);
\draw  (f6) -- (s11);
\draw  (f7) -- (s11);
\draw  (f6) -- (s12);
\draw  (f7) -- (s12);
\draw  (f8) -- (s13);
\draw  (f9) -- (s14);
\draw  (f9) -- (s13);
\draw  (f8) -- (s14);

\end{tikzpicture}}}

\subfigure[Structure of UWGO]{\resizebox{0.4\linewidth}{!}{
   \centering \begin{tikzpicture}[thick,
  bnode/.style={draw,rectangle},
  every fit/.style={ellipse,draw,inner sep=2pt,text width=1cm}]

\begin{scope}[xshift=-2cm, start chain=going right,node distance=1.5cm]
    \node[bnode,on chain] (s1) [] {$v_2$};
    \node[bnode,on chain] (s2)[] {$v_3$};
      
\end{scope}
\node[bnode,xshift=1.2cm] (s5)[] {$v_5$}; 
\node[bnode,xshift=2.5cm] (s7)[] {$v_7$};
\begin{scope}[xshift=-2cm,yshift=-20mm,start chain=going right,node distance=1.5cm]
    \node[bnode,on chain] (s3) [label=above:] {$v_4$};
    \node[bnode,on chain] (s4) [label=above:] {$v_1$};
\end{scope}
\node[bnode,xshift=1.2cm,yshift=-20mm] (s6) [label=above:] {$v_6$};
\node[bnode,xshift=2.3cm,yshift=-20mm,node distance=7mm] (s8) [label=above:] {$v_8$};
\draw (s1) -- (s2) node[midway,above]{$3$};
\draw (s1) -- (s3) node[midway,right]{$4$};
\draw (s1) -- (s4) node[midway,above]{$2$}; 
\draw (s2) -- (s4) node[midway,right]{$3$};
\draw (s2) -- (s3) node[midway,below]{$4$};
\draw (s3) -- (s4) node[midway,above]{$4$};
\draw (s5) -- (s6) node[midway,right]{$2$};
\draw (s7) -- (s6) node[midway,right]{$2$};
\end{tikzpicture}}}
\subfigure[Phases of UWGO vertices]{\resizebox{0.4\linewidth}{!}{
\begin{tikzpicture}
\tkzDefPoints{0/0/P,  2/0/A,  1.5/1.3/B, -0.5/1.93/C, -1.6/-1.2/D, -1.3/-1.5/E, -1/-1.73/F, -.7/-1.87/G}
\tkzDrawCircle(P,A)

\tkzDrawSegment(P,A)
\tkzDrawPoints(P,A)
\tkzLabelPoint[right](A){$\theta_8=0$}

\tkzDrawSegment(P,B)
\tkzDrawPoints(B)
\tkzLabelPoint[right](B){$\theta_5$,$\theta_7=0.23\pi$}

\tkzDrawSegment(P,C)
\tkzDrawPoints(C)
\tkzLabelPoint[above](C){$\theta_6=0.58\pi$}

\tkzDrawSegment(P,D)
\tkzDrawPoints(D)
\tkzLabelPoint[left](D){$\theta_1=1.2\pi$}

\tkzDrawSegment(P,E)
\tkzDrawPoints(E)
\tkzLabelPoint[left](E){$\theta_2=1.27\pi$}

\tkzDrawSegment(P,F)
\tkzDrawPoints(F)
\tkzLabelPoint[left](F){$\theta_3=1.33\pi$}

\tkzDrawSegment(P,G)
\tkzDrawPoints(G)
\tkzLabelPoint[below](G){$\theta_4=1.38\pi$}

\end{tikzpicture}}
}

\caption{Projecting $BG$ to $UWGO$.}\label{fig:projectBGtoUWGO}
\end{figure}       
     For each young butterfly $\bowtie_{v}$, an oscillator vertex $v$ is created with three attributes initialized to zero: phase, frequency, and identifier. The vertex identifier is then set to the hash code\footnote{The Java method hashCode() must consistently return the same integer for `equal' objects during one execution of a Java application. This method is not required to return distinct integers for unequal objects by general contract indicated in \url{https://docs.oracle.com/}. As much as is reasonably practical, the hashCode method defined by class Object does return distinct integers for distinct objects.  Since the UWGO vertex objects are unequal, there is a probability for mapping different vertices to the same ID.} of the object representing $v$ and added to UWGO (Algorithm~\ref{alg:project}, lines~\ref{alg:project:mapstart}-\ref{alg:project:mapend}). The cumulative butterfly count follows a power-law function of the total number of edges (\textit{butterfly densification power-law}~\cite{sheshbolouki2022sgrapp}), therefore we don't use an incremental number for the identifier. Instead we use a fixed-size ID and call the garbage collector afterward (Algorithm~\ref{alg:project}, line~\ref{alg:project:gc}). $v$ is connected to any existing vertex $n$ whose corresponding young butterfly $\bowtie_n$ is in current $W_{BG}$ and shares at least one j-vertex with $\bowtie_{v}$. Alternatively, connections can be based on shared i-vertices. The static weight of the edge between $v$ and $n$ is the number of butterflies adjacent to j-vertices of $\bowtie_{v}$ (Algorithm~\ref{alg:project}, Lines~\ref{alg:project:addedgestart}-\ref{alg:project:addedgeend}). Consequently, the UWGO vertices with higher weighted degrees represent the butterflies which are newer and connected to more and high degree butterflies. The reason is that $W_{BG}$ vertices are stored and iterated in the data structures according to the order of their SGR ingestion/arrival. Therefore, butterflies whose elements are ingested later are identified later and their UWGO vertex is simultaneously connected to the previous ones with an edge weight equal to the number of its UWGO neighbours plus one.  For example in Figure~\ref{fig:projectBGtoUWGO}(b), $v_1$ and $v_2$ are connected by an edge with weight equal to $2$ since $\bowtie_{v_2}$ is identified after $\bowtie_{v_1}$ and they share $j_1$ and $j_2$. Next, $v_3$ is connected to $v_1$ and $v_2$ with weight $3$ and then, $v_4$ is connected to these three vertices with weight $4$. $v_6$ connects to $v_5$, and $v_7$ connects to $v_6$ with weight $2$. 
  
    \textbf{\textcolor{teal}{2.b)}} The local data structures are renewed and garbage collector is called (Algorithm~\ref{alg:project}, lines~\ref{alg:project:setNull}-\ref{alg:project:gc}). Also, all of the SGRs in $W_{BG}$ are retired to avoid redundant updates to the system state (Algorithm~\ref{alg:framework}, line~\ref{alg:cd:BGrenew}). 
   
    \textbf{\textcolor{teal}{2.c)}} The attributes of UWGO vertices are updated  (Algorithm~\ref{alg:framework}, lines~\ref{alg:cd:assignphasefreqstart}-~\ref{alg:cd:assignphasefreqend}). The frequency $\Omega_v$ is sampled from a normal distribution with mean equal to zero and the phase is set as $\theta_v=(\sum_{n\in N(v)} n.ID) \% 2\pi$. 
    The phase of an oscillator embeds the neighbourhood of the corresponding butterfly. Butterflies are the building blocks of the stream. Therefore, projecting $W_{BG}$ to $W_{UWGO}$ implies embedding the stream to a latent space of phases in $[0,2\pi)$.
    In Figure~\ref{fig:projectBGtoUWGO}(c),  $\theta_8=0$ since $\bowtie_{v_8}$ is not connected to other butterflies, $\theta_5=\theta_7$ since $\bowtie_{v_5}$ and $\bowtie_{v_7}$ have one shared neighbour $\bowtie_{v_6}$, and the rest of the vertices except $v_6$ have close phases due to similar neighbourhoods.   

\subsection{SGDD - Drift Detection}\label{sGraddDriftDetection}
The sequential procedures of drift criteria and drift evaluation are as the following (steps 3-6 and yellow boxes in Figure~\ref{fig:proposed}). 

\textbf{Drift Criteria.} A common measure of the level of global phase synchronization in a network of phase oscillators is a quantity called order parameter~\cite{rodrigues2016kuramoto}. We calculate it for sequential instances $W$ of $W_{UWGO}$ with oscillators $V$ as the following.
\begin{align*}
    O[W]=((\sum_{v\in V} \sin \theta_v)^2+(\sum_{v\in V} \cos \theta_v)^2)^{\frac{1}{2}}/|V|
\end{align*} 
 $0\leq O[W]\leq 1$, the higher $O[W]$, the greater the degree of synchronization. $O[W]=1$ denotes a phase synchrony state where all vertices have the same phase, which means butterflies have similar neighbourhoods and are densely connected to each other. Two time-series, $O_1$ and $O_2$,  as the drift criteria are recorded over time:
    
    \textbf{\textcolor{orange!60}{3)}} The order parameter is first computed over $UWGO$'s structure and phases as $O_1[W]$ (Algorithm~\ref{alg:framework}, line~\ref{mainalg:o1}). The phases in Figure~\ref{fig:projectBGtoUWGO}(c) would result in $O_1=0.17$.
   
    \textbf{\textcolor{orange!60}{4)}} Kuramoto model ~\cite{rodrigues2016kuramoto} is the most popular approach to formulate the synchronization process in a population of interacting oscillators. It quantifies the phase change for each oscillator, according to its frequency and the significance of phase difference with its neighbours, such that a global synchronization can be reached. Given the current phases $\Theta=\{\theta_v\}$, edge weights $\{w_{vn}\}$, and frequencies $\{\Omega_v\}$, the phase evolution of a vertex $v$ is denoted as  $\frac{d\theta_v}{dt}= \Omega_v+\sum_{n\in N(v)}w_{vn}sin(\theta_v-\theta_n)$. This ordinary differential equation is solved using the Runge Kutta method with $h=0.01$ (Algorithm~\ref{alg:framework}, line~\ref{mainalg:theta2}).
    \begin{align*}
    \Delta\theta_v  = \frac{h}{6}(K(1)_v+2K(2)_v+2K(3)_v+K(4)_v)\end{align*}
    \begin{align*}K(1)_v  = \Omega_v+\Sigma_{n\in N(v)}w_{vn}\sin{(\theta_n -\theta_v )}\end{align*}
    \begin{align*}K(2)_v  =  \Omega_v+\Sigma_{n\in N(v)}w_{vn}\sin{(\theta_n +\frac{h}{2}K(1)_n-\theta_v -\frac{h}{2}K(1)_v)}\end{align*}
    \begin{align*}K(3)_v  =  \Omega_v+\Sigma_{n\in N(v)}w_{vn}\sin{(\theta_n +\frac{h}{2}K(2)_n-\theta_v -\frac{h}{2}K(2)_v)}\end{align*}
    \begin{align*}K(4)_v  =  \Omega_v+\Sigma_{n\in N(v)}w_{vn}\sin{(\theta_n +hK(3)_n-\theta_v -hK(3)_v)}
    \end{align*}
    The model results in the following phases in Figure~\ref{fig:projectBGtoUWGO}:
    $\Delta\theta_1=0.13$, $\Delta\theta_2=0.01$, $\Delta\theta_3=-0.02$, $\Delta\theta_4=-0.09$, $\Delta\theta_5=0.01$, $\Delta\theta_6=-0.1$, $\Delta\theta_7=0$, $\Delta\theta_8=0.01$.
        
    \textbf{\textcolor{orange!60}{5)}} The order parameter is computed as $O_2[W]$ for predicted phase changes $\Delta\theta$ (Algorithm~\ref{alg:framework}, line~\ref{mainalg:o2}).
    The oscillators in the running example have $O_2=0.98$. 

\textbf{Drift Evaluation.}  The evolutions in $O_1$ and $O_2$ are evaluated to detect a CD (Algorithm~\ref{alg:framework}, line~\ref{mainalg:driftdetect}).
    
    \textbf{\textcolor{orange!60}{6)}} A CD is signaled when the butterfly neighbourhoods are stable, while some butterflies display a future change in their neighbourhoods. Precisely, a drift is signaled when three conditions $C_1$, $C_2$, and $C_3$ are satisfied (Algorithm~\ref{alg:detectdriftv2}).
    \begin{itemize}
        \item [$C_1$:] a local maximum/minimum is observed in $O_2$.
        \item [$C_2$:] $O_1$ remains steadily fixed.
        \item [$C_3$:] The last CD signal has been in at least 10 windows back.
    \end{itemize}
     $C_1$ is implemented as $(10^\alpha\mu_1-10^\alpha O_1[t])/10^\alpha<10^{-\alpha}$, where $\mu_1$ is the average of the last $S^\prime$ values of $O_1$
          and $\alpha$ is a dynamic value used as a threshold and a precision for difference of $\mu_1$ and $O_1[t]$. 
        
        $C_2$ is implemented as ($Ngreater\geq S^\prime$ OR $Nless\geq S^\prime$), where $S^\prime$ is a fraction of $S$ and $Ngreater$ and $Nless$ denote the number of elements in the most recent suffix of $S$ elements in $O_2$ that are greater and less than $O_2[t]$, respectively.
            
        When a drift occurs, the structure of streaming graph perturbs and $O_1$ and $O_2$ experience frequent fluctuations, therefore $C_1$ and $C_2$ should adapt to these perturbations through proper setting of  $S$, $S^\prime$, and $\alpha$. This is to reach a balanced detection state between sensitivity and robustness.
        \begin{itemize}
            \item $S$ is determined based on a function of the number of detections, average burst size, and maximum seen burst size (Algorithm~\ref{alg:detectdriftv2}, line~\ref{v2S}). $S$ fluctuates with changing of burst sizes. when $S$ decreases, $\mu_1$, $Ngreater$, and $Nless$ are computed over smaller suffix to pass the fluctuated values.
            \item  $S^\prime$ decreases as the number of detections increases to ease $C_2$ and avoid missing drifts.
            \item $\alpha$ is calculated as the current number of detections plus two; therefore, as detections increase, $C_1$ gets more difficult to avoid false detections.
        \end{itemize}

   \begin{algorithm}[!ht]\caption{CD Check via Butterflies Interconnections}\label{alg:detectdriftv2}
  \DontPrintSemicolon
   \Fn{\hypertarget{DetectDrift}{$CDC_{Butterfly}$($\Bar{B},maxB,O_1,O_2,t,W,W_dserie$)}}{
    
        $d\gets W_dserie.size()$
        
        $S \gets \frac{\floor{log_{10}(Max(maxB,100))}}{\floor{log_{10}(Max(\Bar{B},10))}}^{(-1)^{d+1}}$ \label{v2S}
        
        $S^\prime \gets (1-d)S $
        
        $\mu_1 \gets$ mean of the last $S^\prime$ values in $O_1$
        
       $Nmore \gets$ number of elements among the last $S$ elements of $O_2$ which are greater than $O_2[W]$
       
        $Nless \gets$ number of elements among the last $S$ elements of $O_2$ which are less than $O_2[W]$
        
        $\alpha\gets d+2$
        
        \If{$(Nless\geq S^\prime \vee  Nmore\geq S^\prime) \bigwedge (|[(10^\alpha\mu_1-10^\alpha O_1[k])]| /10^\alpha)<10^{-\alpha}) \bigwedge W-W_dserie.lastElement()>10$}{
            Signal a drift at sgr index $t$, window $W$, and current system time
            
            $W_dserie.add(W)$
        }
    }
\end{algorithm}
\begin{table*}[!h]
\caption{The average system time distance (ms) \textcolor{cyan}{/} \textcolor{c5}{SGR count distance} of the ith CD and its first and last \emph{SGDP}'s signals  with suffix size determined as $S \gets \frac{\floor{log_{10}(Max(maxB,100))}}{\floor{log_{10}(Max(\Bar{B},10)}})^{(-1)^{d}}$.}
    \centering\resizebox{\textwidth}{!}{
    \begin{tabular}{|c|c|c|c|c|c|c|c|c|c|c|}
        \hline  ms\textcolor{cyan}{/}\textcolor{c5}{sgr}&\cellcolor{yellow}$d_{1f}$&\cellcolor{yellow}$d_{1l}$  &   \cellcolor{orange}$d_{2f}$&\cellcolor{orange}$d_{2l}$    &\cellcolor{mustard}$d_{3f}$&\cellcolor{mustard}$d_{3l}$    &\cellcolor{darkorange}$d_{4f}$&\cellcolor{darkorange}$d_{4l}$ &  \cellcolor{cherrybrown}$d_{5f}$&\cellcolor{cherrybrown}$d_{5l}$ \\\hline\hline

         \cellcolor{green1}$G_{11}$&      621.21\textcolor{cyan}{/}\textcolor{c5}{990}&0.5\textcolor{cyan}{/}\textcolor{c5}{433}&		2863.61\textcolor{cyan}{/}\textcolor{c5}{198998}&110.42\textcolor{cyan}{/}\textcolor{c5}{6914}&		2374.17\textcolor{cyan}{/}\textcolor{c5}{96837}&6.06\textcolor{cyan}{/}\textcolor{c5}{16592}&		36.08\textcolor{cyan}{/}\textcolor{c5}{91028}&5.5415029&		94.71\textcolor{cyan}{/}\textcolor{c5}{14290}&11432.39\textcolor{cyan}{/}\textcolor{c5}{572992}\\
         
         \cellcolor{green1}$G_{12}$&     544.56\textcolor{cyan}{/}\textcolor{c5}{990}&0.54\textcolor{cyan}{/}\textcolor{c5}{433}&		2602.93\textcolor{cyan}{/}\textcolor{c5}{199064}&0.79\textcolor{cyan}{/}\textcolor{c5}{1656}&		2273.68\textcolor{cyan}{/}\textcolor{c5}{89324}&7.67\textcolor{cyan}{/}\textcolor{c5}{20581}&		39.92\textcolor{cyan}{/}\textcolor{c5}{97835}&4.41\textcolor{cyan}{/}\textcolor{c5}{10190}&		4.7\textcolor{cyan}{/}\textcolor{c5}{10631}&6619.2\textcolor{cyan}{/}\textcolor{c5}{516866}\\
         
         \cellcolor{green1}$G_{13}$&     531.91\textcolor{cyan}{/}\textcolor{c5}{990}&0.58\textcolor{cyan}{/}\textcolor{c5}{433}&		1553.38\textcolor{cyan}{/}\textcolor{c5}{199090}&0.49\textcolor{cyan}{/}\textcolor{c5}{866}&		2576.25\textcolor{cyan}{/}\textcolor{c5}{92921}&2.62\textcolor{cyan}{/}\textcolor{c5}{6921}&		40.05\textcolor{cyan}{/}\textcolor{c5}{96075}&2.97\textcolor{cyan}{/}\textcolor{c5}{6728}&		1.56\textcolor{cyan}{/}\textcolor{c5}{3453}&3853.21\textcolor{cyan}{/}\textcolor{c5}{588808}\\
         
         \cellcolor{green1}$G_{14}$&    543.7\textcolor{cyan}{/}\textcolor{c5}{990}&0.533\textcolor{cyan}{/}\textcolor{c5}{433}&		2256.41\textcolor{cyan}{/}\textcolor{c5}{199062}&1.35\textcolor{cyan}{/}\textcolor{c5}{3372}&		1953.98\textcolor{cyan}{/}\textcolor{c5}{95679}&635.61\textcolor{cyan}{/}\textcolor{c5}{6006}&		33.61\textcolor{cyan}{/}\textcolor{c5}{92278}&1.48\textcolor{cyan}{/}\textcolor{c5}{3883}&		4.85\textcolor{cyan}{/}\textcolor{c5}{13423}&3998.81\textcolor{cyan}{/}\textcolor{c5}{569089}\\
         
         \cellcolor{green1}$G_{15}$&    525.28\textcolor{cyan}{/}\textcolor{c5}{990}&0.61\textcolor{cyan}{/}\textcolor{c5}{433}&		3352.88\textcolor{cyan}{/}\textcolor{c5}{199126}&2.74\textcolor{cyan}{/}\textcolor{c5}{7084}&		3128.61\textcolor{cyan}{/}\textcolor{c5}{98075}&4.47\textcolor{cyan}{/}\textcolor{c5}{12774}&		34.06\textcolor{cyan}{/}\textcolor{c5}{89358}&6.05\textcolor{cyan}{/}\textcolor{c5}{15611}&		2.05\textcolor{cyan}{/}\textcolor{c5}{5641}&4326.21\textcolor{cyan}{/}\textcolor{c5}{530006}\\ 
         
         \hline\hline

         \cellcolor{green1}$G_{21}$&   1327.35\textcolor{cyan}{/}\textcolor{c5}{982}&0.49\textcolor{cyan}{/}\textcolor{c5}{433}& &&  					832.89\textcolor{cyan}{/}\textcolor{c5}{16152}&169.31\textcolor{cyan}{/}\textcolor{c5}{4558}&		25834.85\textcolor{cyan}{/}\textcolor{c5}{198742}&83.8\textcolor{cyan}{/}\textcolor{c5}{638}&		157.41\textcolor{cyan}{/}\textcolor{c5}{1843}&25650.69\textcolor{cyan}{/}\textcolor{c5}{397538}\\  
         
         \cellcolor{green1}$G_{22}$&   519.58\textcolor{cyan}{/}\textcolor{c5}{990}&0.48\textcolor{cyan}{/}\textcolor{c5}{433}&		2551.06\textcolor{cyan}{/}\textcolor{c5}{199382}&0.57\textcolor{cyan}{/}\textcolor{c5}{1096}&		2589.36\textcolor{cyan}{/}\textcolor{c5}{185393}&8.26\textcolor{cyan}{/}\textcolor{c5}{24384}&		75.53\textcolor{cyan}{/}\textcolor{c5}{194324}&5.8\textcolor{cyan}{/}\textcolor{c5}{14140}&		5.7\textcolor{cyan}{/}\textcolor{c5}{12664}&3187.11\textcolor{cyan}{/}\textcolor{c5}{394536}\\  
         
         \cellcolor{green1}$G_{23}$&     574.21\textcolor{cyan}{/}\textcolor{c5}{990}&0.56\textcolor{cyan}{/}\textcolor{c5}{433}&		3118.22\textcolor{cyan}{/}\textcolor{c5}{199220}&1.12\textcolor{cyan}{/}\textcolor{c5}{3069}&		5497.82\textcolor{cyan}{/}\textcolor{c5}{184920}&5.55\textcolor{cyan}{/}\textcolor{c5}{16546}&		58.85\textcolor{cyan}{/}\textcolor{c5}{179858}&1.47\textcolor{cyan}{/}\textcolor{c5}{3710}&		25.7\textcolor{cyan}{/}\textcolor{c5}{68088}&377.27\textcolor{cyan}{/}\textcolor{c5}{359944}\\
         
         \cellcolor{green1}$G_{24}$&   522.79\textcolor{cyan}{/}\textcolor{c5}{990}&0.5\textcolor{cyan}{/}\textcolor{c5}{433}&		2907.32\textcolor{cyan}{/}\textcolor{c5}{199156}&0.3\textcolor{cyan}{/}\textcolor{c5}{733}&		3199.19\textcolor{cyan}{/}\textcolor{c5}{184041}&636.2\textcolor{cyan}{/}\textcolor{c5}{36446}&		74.45\textcolor{cyan}{/}\textcolor{c5}{196190}&7.2\textcolor{cyan}{/}\textcolor{c5}{17563}&		6.22\textcolor{cyan}{/}\textcolor{c5}{15330}&3290.17\textcolor{cyan}{/}\textcolor{c5}{349942}\\  
         
         \cellcolor{green1}$G_{25}$&     523\textcolor{cyan}{/}\textcolor{c5}{990}&0.48\textcolor{cyan}{/}\textcolor{c5}{433}&		3053.84\textcolor{cyan}{/}\textcolor{c5}{199001}&0.78\textcolor{cyan}{/}\textcolor{c5}{1652}&		1552.73\textcolor{cyan}{/}\textcolor{c5}{181108}&17.18\textcolor{cyan}{/}\textcolor{c5}{49227}&		73.49\textcolor{cyan}{/}\textcolor{c5}{198918}&10.45\textcolor{cyan}{/}\textcolor{c5}{25915}&		1.7\textcolor{cyan}{/}\textcolor{c5}{4192}&3981.72\textcolor{cyan}{/}\textcolor{c5}{307993}\\
         \hline
         \tiny AVG& 623.3\textcolor{cyan}{/}\textcolor{c5}{989.2} &0.5\textcolor{cyan}{/}\textcolor{c5}{433}  &2695.5\textcolor{cyan}{/}\textcolor{c5}{199122.1} &13.2\textcolor{cyan}{/}\textcolor{c5}{2938}&    2597.9\textcolor{cyan}{/}\textcolor{c5}{122445} &149.3\textcolor{cyan}{/}\textcolor{c5}{19403.5}&
             2630.1\textcolor{cyan}{/}\textcolor{c5}{143460.6}& 12.9\textcolor{cyan}{/}\textcolor{c5}{11340.7}&  30.5\textcolor{cyan}{/}\textcolor{c5}{14955.5}& 6671.7\textcolor{cyan}{/}\textcolor{c5}{458771.4} \\
         \hline
         
         \hline
    \end{tabular}}
    
    \centering\resizebox{\textwidth}{!}{
    \begin{tabular}{|c|c|c|c|c|c|c|c|c|}
        \hline  ms\textcolor{cyan}{/}\textcolor{c5}{sgr}& \cellcolor{yellow}$d_{1f}$&\cellcolor{yellow}$d_{1l}$  &   \cellcolor{orange}$d_{2f}$&\cellcolor{orange}$d_{2l}$    &\cellcolor{mustard}$d_{3f}$&\cellcolor{mustard}$d_{3l}$    &\cellcolor{darkorange}$d_{4f}$&\cellcolor{darkorange}$d_{4l}$ \\\hline\hline
        
        \cellcolor{lightpurple}$R_{11}$&  552.8\textcolor{cyan}{/}\textcolor{c5}{990}&0.5\textcolor{cyan}{/}\textcolor{c5}{433}&     2366.4\textcolor{cyan}{/}\textcolor{c5}{199007}&1.82\textcolor{cyan}{/}\textcolor{c5}{4682}&		606.01\textcolor{cyan}{/}\textcolor{c5}{92643}&11.05\textcolor{cyan}{/}\textcolor{c5}{27346}&     0.28\textcolor{cyan}{/}\textcolor{c5}{468}&291.62\textcolor{cyan}{/}\textcolor{c5}{683795}  \\
         
        \cellcolor{lightpurple}$R_{12}$&  527.08\textcolor{cyan}{/}\textcolor{c5}{990}&0.44\textcolor{cyan}{/}\textcolor{c5}{433}&   2315.66\textcolor{cyan}{/}\textcolor{c5}{199004}&1.4\textcolor{cyan}{/}\textcolor{c5}{3177}&       940\textcolor{cyan}{/}\textcolor{c5}{93872}&1.35\textcolor{cyan}{/}\textcolor{c5}{3458}&		    3.29\textcolor{cyan}{/}\textcolor{c5}{9926}&268.4\textcolor{cyan}{/}\textcolor{c5}{685600}\\
         
        \cellcolor{lightpurple}$R_{13}$&  505.67\textcolor{cyan}{/}\textcolor{c5}{990}&0.45\textcolor{cyan}{/}\textcolor{c5}{433}&   2497.91\textcolor{cyan}{/}\textcolor{c5}{199159}&2.99\textcolor{cyan}{/}\textcolor{c5}{6894}&		3930.04\textcolor{cyan}{/}\textcolor{c5}{99564}&6.09\textcolor{cyan}{/}\textcolor{c5}{18604}&		1.61\textcolor{cyan}{/}\textcolor{c5}{4433}&274.23\textcolor{cyan}{/}\textcolor{c5}{698234} \\
         
        \cellcolor{lightpurple}$R_{14}$&  526.19\textcolor{cyan}{/}\textcolor{c5}{990}&0.43\textcolor{cyan}{/}\textcolor{c5}{433}&	2616.27\textcolor{cyan}{/}\textcolor{c5}{199090}&1.32\textcolor{cyan}{/}\textcolor{c5}{3252}&		3397.48\textcolor{cyan}{/}\textcolor{c5}{92309}&3.81\textcolor{cyan}{/}\textcolor{c5}{10564}&		2.94\textcolor{cyan}{/}\textcolor{c5}{8373}&276.94\textcolor{cyan}{/}\textcolor{c5}{691594} \\

        \cellcolor{lightpurple}$R_{15}$&  503.83\textcolor{cyan}{/}\textcolor{c5}{990}&0.43\textcolor{cyan}{/}\textcolor{c5}{433}&    3592.33\textcolor{cyan}{/}\textcolor{c5}{199040}&1.63\textcolor{cyan}{/}\textcolor{c5}{4427}&		2384.35\textcolor{cyan}{/}\textcolor{c5}{85237}&944.28\textcolor{cyan}{/}\textcolor{c5}{11098}&	5.16\textcolor{cyan}{/}\textcolor{c5}{16380}&243.57\textcolor{cyan}{/}\textcolor{c5}{670088}  \\

         \hline\hline
         
         \cellcolor{lightpurple}$R_{21}$&   633.1\textcolor{cyan}{/}\textcolor{c5}{990}&0.49\textcolor{cyan}{/}\textcolor{c5}{433}&		4665.46\textcolor{cyan}{/}\textcolor{c5}{399216}&5.1\textcolor{cyan}{/}\textcolor{c5}{13644}&		6762.54\textcolor{cyan}{/}\textcolor{c5}{195310}&20.54\textcolor{cyan}{/}\textcolor{c5}{56107}&		1.75\textcolor{cyan}{/}\textcolor{c5}{3939}&145.13\textcolor{cyan}{/}\textcolor{c5}{369707}\\
         
         \cellcolor{lightpurple}$R_{22}$&   601.17\textcolor{cyan}{/}\textcolor{c5}{990}&0.56\textcolor{cyan}{/}\textcolor{c5}{433}&		3424.14\textcolor{cyan}{/}\textcolor{c5}{399050}&0.4\textcolor{cyan}{/}\textcolor{c5}{}896&		6018.52\textcolor{cyan}{/}\textcolor{c5}{184303}&9.72\textcolor{cyan}{/}\textcolor{c5}{26277}&		2.68\textcolor{cyan}{/}\textcolor{c5}{6651}&167.59\textcolor{cyan}{/}\textcolor{c5}{398792}\\
         
         \cellcolor{lightpurple}$R_{23}$&   557.2\textcolor{cyan}{/}\textcolor{c5}{990}&0.55\textcolor{cyan}{/}\textcolor{c5}{433}&		3424.61\textcolor{cyan}{/}\textcolor{c5}{399183}&2.59\textcolor{cyan}{/}\textcolor{c5}{7105}&		3534.66\textcolor{cyan}{/}\textcolor{c5}{192765}&5.52\textcolor{cyan}{/}\textcolor{c5}{16287}&		4.02\textcolor{cyan}{/}\textcolor{c5}{11450}&162.62\textcolor{cyan}{/}\textcolor{c5}{389676}\\
         
         \cellcolor{lightpurple}$R_{24}$&   550.89\textcolor{cyan}{/}\textcolor{c5}{990}&0.48\textcolor{cyan}{/}\textcolor{c5}{433}&		2609.36\textcolor{cyan}{/}\textcolor{c5}{399183}&0.46\textcolor{cyan}{/}\textcolor{c5}{1053}&		3588.61\textcolor{cyan}{/}\textcolor{c5}{189434}&5.58\textcolor{cyan}{/}\textcolor{c5}{13771}&		2.52\textcolor{cyan}{/}\textcolor{c5}{6104}&174.47\textcolor{cyan}{/}\textcolor{c5}{390063}\\
         
         \cellcolor{lightpurple}$R_{25}$&   522.73\textcolor{cyan}{/}\textcolor{c5}{990}&0.5\textcolor{cyan}{/}\textcolor{c5}{433}&		3695.39\textcolor{cyan}{/}\textcolor{c5}{399188}&3.15\textcolor{cyan}{/}\textcolor{c5}{8308}&		3286.01\textcolor{cyan}{/}\textcolor{c5}{195456}&1.92\textcolor{cyan}{/}\textcolor{c5}{5126}&		5.61\textcolor{cyan}{/}\textcolor{c5}{16303}&157.09\textcolor{cyan}{/}\textcolor{c5}{380914}\\
         \hline
         \tiny AVG& 548.1\textcolor{cyan}{/}\textcolor{c5}{990} &0.5 \textcolor{cyan}{/}\textcolor{c5}{433}  &3120.7\textcolor{cyan}{/}\textcolor{c5}{310222} &2.1\textcolor{cyan}{/}\textcolor{c5}{5416.8}&    3444.8\textcolor{cyan}{/}\textcolor{c5}{142089.3} &101\textcolor{cyan}{/}\textcolor{c5}{18863.8}&
             3\textcolor{cyan}{/}\textcolor{c5}{8402.7} &216.2\textcolor{cyan}{/}\textcolor{c5}{535846.3} \\
         \hline
    \end{tabular}}
    
    \label{tab:SGDP0}
\end{table*}

\end{document}